\documentclass[12pt]{article}
\usepackage[left=1.1in,right=1in,top=1in,bottom=1.3in]{geometry}
\usepackage{float}
\usepackage{booktabs,color}
\usepackage{multirow}
\usepackage{amsmath}
\usepackage{subcaption}
\usepackage{longtable}
\usepackage{adjustbox}
\usepackage{amsfonts}
\usepackage[euler]{textgreek}
\usepackage{enumerate}
\usepackage{textcomp}
\usepackage{graphicx}
\usepackage[outdir=./]{epstopdf}
\usepackage{natbib}

\usepackage{soul} 

\numberwithin{equation}{section}

\usepackage{setspace}
\begin{document}

\title{Asynchronous ADRs: Overnight vs Intraday Returns and Trading Strategies\thanks{This project is partially supported by the 2016 Dean's Office Summer Research Fund provided by the Fu Foundation School of Engineering at Columbia University.}}
\author{Tim Leung\thanks{  {Department of Applied Mathematics, Computational Finance and Risk Management Program, University of Washington, Seattle WA 98195. E-mail:
\mbox{timleung@uw.edu}. Corresponding author. }} \and Jamie Kang\thanks{ {Industrial
Engineering \& Operations Research Department, Columbia University,
New York, NY 10027.  E-mail:
\mbox{jk3667@columbia.edu}.  }} }\date{\today} \maketitle

\begin{abstract} 
American Depositary Receipts (ADRs) are exchange-traded certificates that represent shares of non-U.S. company securities. They are major financial instruments for investing in foreign companies. Focusing on Asian ADRs  in the context of asynchronous markets, we present methodologies and results of empirical analysis of their returns. In particular, we dissect  their returns into intraday and overnight components with respect to the U.S. market hours. The return difference between the S\&P500 index, traded through the SPDR S\&P500 ETF (SPY), and each ADR is found to be a mean-reverting time series, and is fitted to an  Ornstein-Uhlenbeck process  via    maximum-likelihood estimation (MLE). Our empirical observations also lead us to develop and backtest  pairs trading strategies to exploit the   mean-reverting ADR-SPY spreads.  We find consistent positive payoffs when long position in ADR and short position in SPY are simultaneously executed at selected  entry and exit levels.  
\end{abstract}

\newpage 
\section{Introduction}\label{sect-intro}

American Depositary Receipts (ADRs) are certificates, denominated in U.S. dollars,  that represent shares of non-U.S. company securities. There are about 2000 ADRs      traded on U.S.   exchanges, namely the New York Stock Exchange (NYSE) and   NASDAQ, or through the over-the-counter (OTC) market, representing   shares of companies from at least 70 different countries.  ADRs  are among the most direct and popular financial  instruments for investing  in foreign companies, and  have been actively used to diversify portfolios.\footnote{Source:  \emph{Investor Bulletin: American Depositary Receipts}, published by the U.S. Securities and Exchange Commission (SEC)}  In 2015, foreign equity holdings -- through ADRs and local shares -- accounted for 19\% of U.S. investors' equity portfolios. Specifically, total global investments in DRs, both American and non-American, were estimated to be approximately \$1 trillion USD with about 90\% specifically in ADRs according to reports by JP Morgan.\footnote{Source:  \textit{Depositary Receipts Year in Reviews} in 2011 and 2015, published by J.P.Morgan \& Co.} Some of the most traded Exchange-Traded Funds (ETFs) are comprised of ADRs as well. For instance, 14 Chinese ADRs were added to the MSCI Emerging Market Index   in late 2015. The  iShares MSCI Emerging Markets ETF (EEM), which tracks this index, has a market capitalization of \$24 billion  and   daily average volume of 72 million shares.

Due to  their cross-border nature, ADRs' returns are significantly dependent on the market sentiments of both the U.S. and originating home markets. Especially, ADRs from Asian countries, such as Japan, Hong Kong, China, Taiwan, India, and Korea, are of particular interest for two reasons. First, these ADRs  provide U.S. investors with direct  viable   means to invest   in foreign and emerging markets. Consequently, Asian DRs listed on non-Asian exchanges accounted for over 44\% of the total DR market capitalization by the third quarter of 2015.\footnote{Source:  \textit{Depositary Receipts Year in Reviews} in 2011 and 2015, published by J.P.Morgan \& Co.}

 Second, the time zones for the aforementioned countries are 13 to 14 hours ahead of New York's Eastern Standard Time (EST). Therefore, the trading hours of the U.S. market and their home markets do not overlap. The originating home markets open after the U.S. market closes and vice versa. This leads to not only their asynchronous returns with underlying equities but also the   price fluctuations during and after the U.S. market hours. The connection (or contrast) between  completely asynchronous markets has been of particular interest to especially multi-asset investors and other institutional investors. \cite{bergomi2010correlations} provides an interesting insight into the correlations in asynchronous markets by using empirical data of the Stoxx50, S\&P500, and Nikkei indexes,  and applies it to the case of correlation swaps. Similarly, the time zone discrepancy allows us to split their returns into intraday and overnight components, each with fundamentally  different price driving factors.  The method of separating   daily returns into intraday and overnight returns has been applied and anlayzed in previous studies. For instance,  \mbox{\cite{riedel2015risk}} examine and contrast  the risk profiles of intraday and overnight returns of major stock indexes, and find that    overnight returns have lower volatility  but  are subject to much heavier unpredictable innovation tail risks.  For the Asian ADRs studied herein, intraday returns can be intuitively attributed to the news, conditions, and outlook of the U.S. market, whereas  overnight returns are predominantly driven by  the local Asian markets. We seek to understand the volatility contribution by different effects from each market.

In this paper, we analyze a number of  characteristics of ADRs, including  the key statistics of the   intraday returns and overnight returns, their  distributions  and  correlations with the U.S. market. In particular, the correlation of each ADR with respect to the U.S. market is useful  since it can serve as the basis of some trading strategies. Since intraday returns reflect price fluctuations during the U.S. market hours, they are expected to have a relatively high correlation with the U.S. market returns. On a similar note, overnight returns are expected to have a significantly lower correlation. In our  study, we determine  these correlations,  quantitatively contrast their effects on the overall volatility and performance of each ADR, and apply our findings to develop  pair trading strategies. 

Existing literature on ADRs primarily focuses on the relationships between ADRs and their underlying stocks. \cite{patel2015adrs} finds their long-run equilibrium relationship, proposing that ADRs tend to lead underlying stocks as ADR markets react faster to asymmetric information. When it comes to cross-listed equities, studies on the degree of market integration have been done previously. For instance, \cite{koulakiotis2010impact} examine the impact of cross-listings and exchange rate differences on the UK and German markets integration in terms of volatility transmissions and persistence. \cite{madan} studies the correlations between the local stock and the
foreign currency price of the US dollar through the prices of ADR options.   \cite{cai2011pricing} study the co-integration between cross-listed Chinese A- and H-shares prices in terms of differential market sentiments as well as other macro and micro factors. We extend this by investigating the dynamics of ADR pricing with regards to the U.S. market sentiment.  {A recent study     by \mbox{\cite{rodriguez2015chinese}} focuses on single-listed Chinese ADRs, and finds  significant correlations  with the S\&P500 index. Another interesting note regarding price discrepancies between ADRs and underlying stocks has been suggested by \cite{eichler2009adr}. Their price spreads are used as earlier indicators for market risks, specifically currency crisis. Price discrepancies between foreign underlying equities and their U.S. trackers can also  be examined with regards to currency exposure. \cite{williams2016currency} investigates the effect of the underlying foreign currency volatility on country ETFs and proposes an optimal hedge ratio to mitigate the risk. \cite{lee2015industry} identify some  financial and fundamental  factors  that affect the co-movements of ADRs with their home market or the U.S. market. Similarly,  \mbox{\cite{gupta2016linkages}} point to the close long-term relationship between the returns of emerging market ADRs and their home countries' macroeconomic indicators. Some previous studies also adopt the method of separation of ADRs' returns into intraday and overnight components. \cite{he2012day}, for example, reach a conclusion that Chinese ADRs' day returns are most significantly affected by the U.S. market, which agrees with our findings in this paper.

Nonetheless, very few studies investigate the stochastic behavior of return spreads between ADRs and SPY (as a proxy of the U.S. market) despite its potential applications. And thus, in this paper, we select 15 of the most traded ADRs from 5 Asian nations (Japan, Korea, Hong Kong/China, India, Taiwan) and examine the return spreads' mean-reverting process in terms of their different time components. We first compute and compare the means and standard deviations of intraday and overnight ADR returns. We find that ADRs' overnight returns are significantly large in magnitude, if not larger than intraday returns as shown in the cases of several Chinese ADRs. Using QQplots, we then examine the normality of ADR returns to test the conventional notion of normally distributed stock returns. The returns (both intraday and overnight) are slightly heavy-tailed, which can be attributed to factors including occasional extreme price changes.

Among the 15 ADRs considered in our study, the intraday returns of  14 ADRs  show higher correlation with the U.S. market than overnight returns do, for which we use the SPDR S\&P500 ETF (SPY) as a proxy of the U.S. market. This confirms the initial theory that ADRs' intraday returns are driven by the U.S. market sentiments and overnight returns by home market sentiments. We then investigate the stochastic behavior of the discrepancy between ADRs and the U.S. market. This is executed by analyzing the time series of return spreads between ADRs and SPY to capture the mean-reverting process. Some ADR-SPY pairs are chosen to be modeled as Ornstein-Uhlenbeck (OU) processes.  We then use maximum likelihood estimation   (MLE) method to determine the  parameters of the OU models. Lastly, pairs trading strategies are constructed based on the findings. For related  studies on pairs trading of  ETFs, we refer to \cite{meanreversionbook2016,LeungSantoli2016ETF}, and references therein.

The rest of the paper is structured as follows. In Section \ref{sec2}, we analyze the  intraday, overnight, and daily historical returns of ADRs. Then we compare the price dynamics of ADRs and the S\&P500 index in Section \ref{sec3}. Motivated by the empirical findings, we develop in Section \ref{sec4} pairs trading strategies based on the mean-reverting ADR-SPY spreads. Finally, we conclude our study in  Section \ref{conclu}.

\section{Analysis of ADR Returns }\label{sec2}
For our  analysis of intraday and overnight returns, we establish the following criteria for our    ADR selection.
\begin{itemize}
\item Their originating home markets must be in the opposite time zone (relative to the Eastern Standard Time Zone).
\item They must have a substantial amount of market capitalization. Hence, we select those with the largest market capitalization.
\end{itemize}
As such, three  ADRs from different countries, China, Japan, South Korea, Taiwan, and India, are chosen accordingly as listed in Table \ref{tab:ADRlist}. ADRs from other areas, such as Europe, are excluded for the purpose of time-zone comparisons.


We begin by separating the returns into intraday  and overnight components. By doing so, we attribute intraday price fluctuations mainly to the U.S. markets, and those during overnight to other factors, such as home market sentiments.

Using the prices available on  Wharton Research Data Services (WRDS) and Yahoo Finance, we collect the daily opening prices, $P^{open}_{i}$, and closing prices, $P^{close}_{i}$, from 6/15/2004 to 6/13/2014 (2517 trading days, 10 years) for every sample ADR. We further adjust the closing prices to neutralize any effect of dividends or stock splits. These price data are then used to calculate intraday, overnight, and daily returns by:
 \begin{align}
R_{ID}(i) = \frac{P^{close}_i-P^{open}_i}{P^{open}_i}\,, \\
R_{ON}(i) = \frac{P^{open}_i-P^{close}_{i-1}}{P^{close}_{i-1}} \,,\\
R_{DD}(i) = \frac{P^{close}_i-P^{close}_{i-1}}{P^{close}_{i-1}}\,,
\label{eq:Return}
\end{align}
for  $i \in \{1,\dots,2516\}$,  where $R_{ID}$, $R_{ON}$, $R_{DD}$ represent intraday, overnight, and daily returns respectively.

 The unique opposite timezone characteristics of the selected ADRs give rise to our initial hypothesis that overnight returns are the dominant driving force of the overall ADR price fluctuations. Hence, overnight returns are expected to have relatively greater magnitudes, and thus higher standard deviations. One idea that leads to this hypothesis is that, during U.S. market hours (i.e. intraday), ADR prices tend to fluctuate gradually in response to the U.S. market which SPY (SPDR S\&P 500 ETF) closely tracks. However, the degree of such fluctuation is expected to be smaller, yet not negligible, than overnight price fluctuations that come from foreign factors when their home markets are open. To confirm this, we first examine the basic statistics of ADR overnight, intraday, and daily return time series.

\vspace{15pt}
\begin{table}[H]
\begin{center}
\begin{tabular}{l|ccc}
\toprule
Country & Name & Ticker & Mkt.Cap. (\$bil)\\
\toprule
\multirow{3}{*}{China} & China Mobile & CHL & 266 \\
& Petro China & PTR & 343 \\
& China Life Insurance & LFC & 151 \\
\midrule
\multirow{3}{*}{Japan} & Toyota Motor & TM & 230 \\
& Mitsubishi UFJ Financial & MTU & 99 \\
& Nippon Telegraph and Telephone& NTT & 80\\
\midrule
\multirow{3}{*}{South Korea} & Korea Electric Power & KEP & 24 \\
& SK Telecom & SKM & 18\\
& Shinhan Financial & SHG & 17 \\
\midrule
\multirow{3}{*}{Taiwan} & Taiwan Semiconductor & TSM & 119 \\
& Chunghwa Telecom & CHT & 24 \\
& Advanced Semiconductor Engineering & ASX & 10 \\
\midrule
\multirow{3}{*}{India} & HDFC Bank & HDB & 39 \\
& Infosys Technology & INFY & 35 \\
& Wipro & WIT & 21\\
\bottomrule
\end{tabular}
\caption{\small{Asian ADRs with names, tickers, and market capitalizations from  5 Asian countries. ADRs with the highest market cap. in their respective countries are chosen.}}
\label{tab:ADRlist}
\end{center}
\end{table}


\newpage
\subsection{Comparison of intraday, overnight, and daily returns}\label{sec2.1}
We conjecture that a significant portion of ADR returns is realized  after the U.S. market hours (i.e. overnight). To better understand this, we quantitatively analyze the selected ADRs in terms of the means and standard deviations of their intraday, overnight and daily returns. These results are reported in Table \ref{tab:returnstat}. We utilize the return time series, $R_{ON}(i)$ (overnight return), $R_{ID}(i)$ (intraday return), and $R_{DD}(i)$ (daily return), that have been previously constructed using Equation \eqref{eq:Return} based on empirical data. Specifically, we use historical prices from June 2004 to June 2014, a total of 2517 trading days and thus 2516 returns. 
Additionally, we compare the daily proportion of overnight returns in terms of the \emph{squared returns ratio} defined by 
\begin{align}
q(i) &= \frac{R_{ON}^2(i)}{R_{ID}^2(i)+R_{ON}^2(i)}\,,  \label{eq:q}
\end{align}
and aggregate to get 
\begin{align}
\bar{q} &= \frac{1}{N}\sum_{i=1}^{N} q(i)\,,
\label{eq:bar-q}
\end{align}
where $i \in \{1,\ldots, N\}$ and $N = 2516$. From the defining equation \eqref{eq:q}
, the daily proportion $q(i)$ lies between 0 and 1. A higher value of $q(i)$ implies a higher   contribution  to the daily return  by overnight return than by the intraday return.

In Table \ref{tab:returnstat}, we show that  overnight returns have  higher standard deviations than intraday returns do for most ADRs, except in the cases of the three Indian ADRs (HDB, INFY, WIT) and one Taiwanese ADR (TSM). This confirms that most significant (in terms of magnitudes) price fluctuations tend to  occur during overnight periods. There are no observable trends in mean, as returns fluctuate across negative and positive values, canceling out each value and ultimately reaching near zero. Hence, rather than analyzing the means, it seems more reasonable to observe the standard deviations instead. Furthermore, the average ratio  $\bar{q}>0.5$ for the ADRs from China, Japan, and Korea, unlike SPY's average ratio  $\bar{q}=0.39$, corroborating again that these ADR prices fluctuate more heavily after U.S. market hours. Other countries, India and Taiwan, seem to deviate from this claim as they did earlier in standard deviations. Also note that unlike the ADRs, SPY has the intraday return standard deviation much higher than that of the overnight return, which is self-evident given that SPY's underlying market is in the United States, and thus it is most affected by the U.S. market sentiments. 

We show  in Figure \ref{fig:Ratio} another notable observation with regards to the ratio $q$. For most ADRs, the distribution  of the ratio, $q$, is spread out between $[0,1]$, rather than being centralized around $\bar{q}$, thereby resembling a U-shape. However, exceptions occur for 2 Japanese, 2 Korean, and all of Taiwanese ADRs, where we observe that the ratio  peaks around 0.5, giving rise to a W-shape distribution. In particular,  both MTU and ASX display two other smaller peaks around the values 0.2 and 0.8, as  illustrated in Figure \ref{fig:Ratio} (see panels (e) and (f)). \\

\begin{table}[t]
\begin{small}
 \centering
 \begin{tabular}{ll|  cc  cc  cc  |c}
    \toprule
          &       & \multicolumn{2}{c}{\textbf{Overnight}} & \multicolumn{2}{c}{\textbf{Intraday}} & \multicolumn{2}{c}{\textbf{Daily}} & \multicolumn{1}{|c}{\multirow{2}[0]{*}{}} \\
                    &       & \multicolumn{2}{c}{\textbf{$R_{ON}$}} & \multicolumn{2}{c}{\textbf{$R_{ID}$}} & \multicolumn{2}{c}{\textbf{$R_{DD}$}} & \multicolumn{1}{|c}{} \\
          \midrule
Country & ADR & $\mu (10^{-2}\%)$  & $\sigma (\%)$  & $\mu (10^{-2}\%)$  & $\sigma (\%)$  & $\mu (10^{-2}\%)$  & $\sigma (\%)$  & $\bar{q}$\\
  \toprule
\multicolumn{1}{c}{\multirow{3}[0]{*}{\textbf{CHINA}}} & \multicolumn{1}{c|}{\textbf{CHL}} & 3.79                           & 1.60      & 3.09                           & 1.29      & 7.13                           & 2.17      & 0.56   \\
\multicolumn{1}{c}{}                                & \multicolumn{1}{c|}{\textbf{LFC}}  & 1.84                           & 1.97      & 7.40                           & 1.49      & 9.55                           & 2.61      & 0.57   \\
\multicolumn{1}{c}{}                                & \multicolumn{1}{c|}{\textbf{PTR}}  & 3.17                           & 1.71      & 3.23                           & 1.45      & 6.71                           & 2.38      & 0.52   \\
        \midrule
\multicolumn{1}{c}{\multirow{3}[0]{*}{\textbf{INDIA}}} & \multicolumn{1}{c|}{\textbf{HDB}}  & 8.02                           & 1.74      & 4.09                           & 2.16      & 12.39                          & 2.88      & 0.41   \\
\multicolumn{1}{c}{}                                & \multicolumn{1}{c|}{\textbf{INFY}} & 0.75                           & 1.59      & 5.61                           & 1.69      & 6.56                           & 2.41      & 0.43   \\
\multicolumn{1}{c}{}                                & \multicolumn{1}{c|}{\textbf{WIT}} & 4.70                           & 1.71      & 3.28                           & 2.36      & 7.43                           & 2.74      & 0.39   \\        
\midrule
\multicolumn{1}{c}{\multirow{3}[0]{*}{\textbf{JAPAN}}}          & \multicolumn{1}{c|}{\textbf{MTU}}           & -9.94                          & 1.84      & 11.46                          & 1.56      & 1.22                           & 2.29      & 0.55   \\
\multicolumn{1}{c}{}                                & \multicolumn{1}{c|}{\textbf{NTT}}                      & -5.96                          & 1.35      & 8.04                           & 0.90      & 2.02                           & 1.60      & 0.60   \\
\multicolumn{1}{c}{}                                & \multicolumn{1}{c|}{\textbf{TM}}                       & 3.87                           & 1.34      & -0.92                          & 1.03      & 3.01                           & 1.72      & 0.57   \\
\midrule
\multicolumn{1}{c}{\multirow{3}[0]{*}{\textbf{KOREA}}}            & \multicolumn{1}{c|}{\textbf{KEP}}                      & 4.09                           & 1.80      & 1.69                           & 1.60      & 6.02                           & 2.52      & 0.52   \\
                                                    & \multicolumn{1}{c|}{\textbf{SHG}}                      & 1.64                           & 2.23      & 6.84                           & 1.73      & 8.91                           & 2.99      & 0.57   \\
                                                    & \multicolumn{1}{c|}{\textbf{SKM}}                      & -0.19                          & 1.36      & 2.63                           & 1.28      & 2.25                           & 1.77      & 0.49   \\
\midrule
\multicolumn{1}{c}{\multirow{3}[0]{*}{\textbf{TAIWAN}}}                    & \multicolumn{1}{c|}{\textbf{ASX}}                      & 5.62                           & 1.98      & 0.48                           & 1.92      & 6.06                           & 2.74      & 0.47   \\
                                                    & \multicolumn{1}{c|}{\textbf{CHT}}                      & -5.69                          & 1.36      & 9.69                           & 1.25      & 3.97                           & 1.80      & 0.45   \\
                                                    & \multicolumn{1}{c|}{\textbf{TSM}}                      & 7.53                           & 1.27      & -1.23                          & 1.75      & 6.30                           & 2.16      & 0.41   \\
        \midrule
\multicolumn{2}{c|}{\textbf{SPY}}                                                    & 2.09                           & 0.70      & 0.79                           & 1.02      & 2.93                           & 1.28      & 0.39  \\
   \bottomrule
   \end{tabular}
  \caption{\small{Statistics of intraday, overnight, and daily returns using empirical price data (with adjusted close) during 6/15/2004--6/13/2014. For every return time series, $\mu$ = mean and $\sigma$ = standard deviation are reported. Lastly, the average ratios  $\bar{q}$ (see \eqref{eq:bar-q})  are  recorded in the last column. For comparison, we also report the same statistics of SPY.}}
   \label{tab:returnstat}
  \end{small}
\end{table}

\begin{figure}[H]
\centering
\begin{subfigure}[b]{0.49\textwidth}
\includegraphics[width=\textwidth]{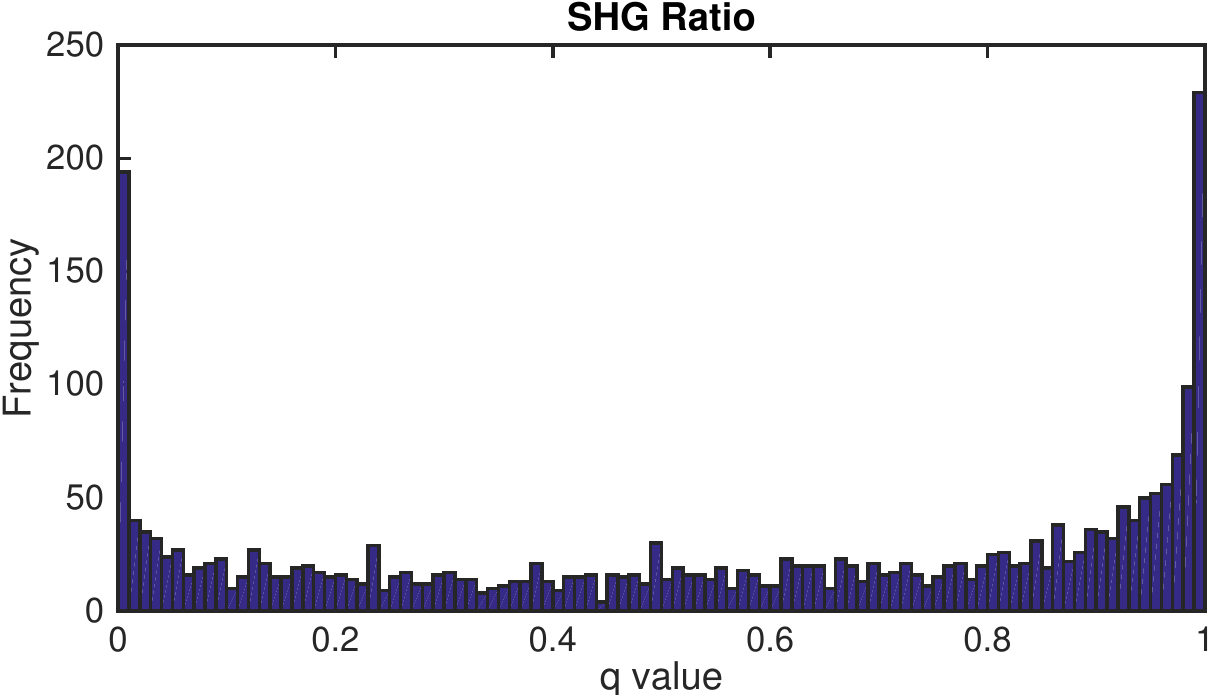}
\caption{ }
\end{subfigure}
\begin{subfigure}[b]{0.49\textwidth}
\includegraphics[width=\textwidth]{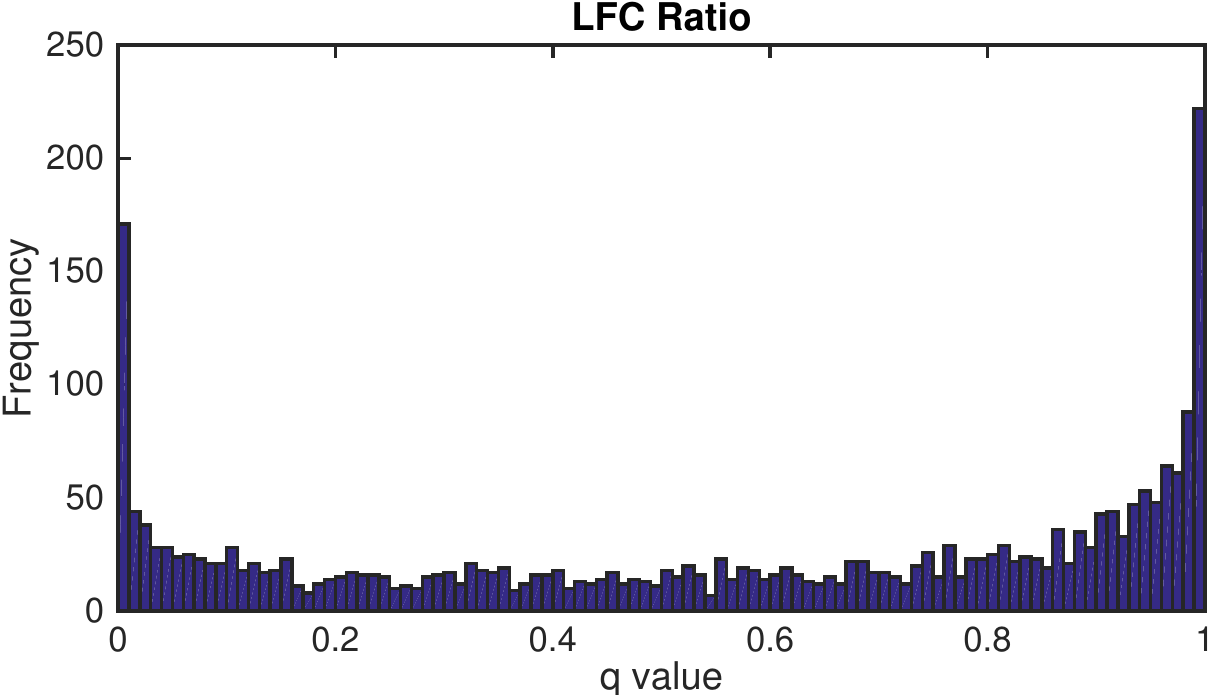}
\caption{ }
\end{subfigure}
\begin{subfigure}[b]{0.49\textwidth}
\includegraphics[width=\textwidth]{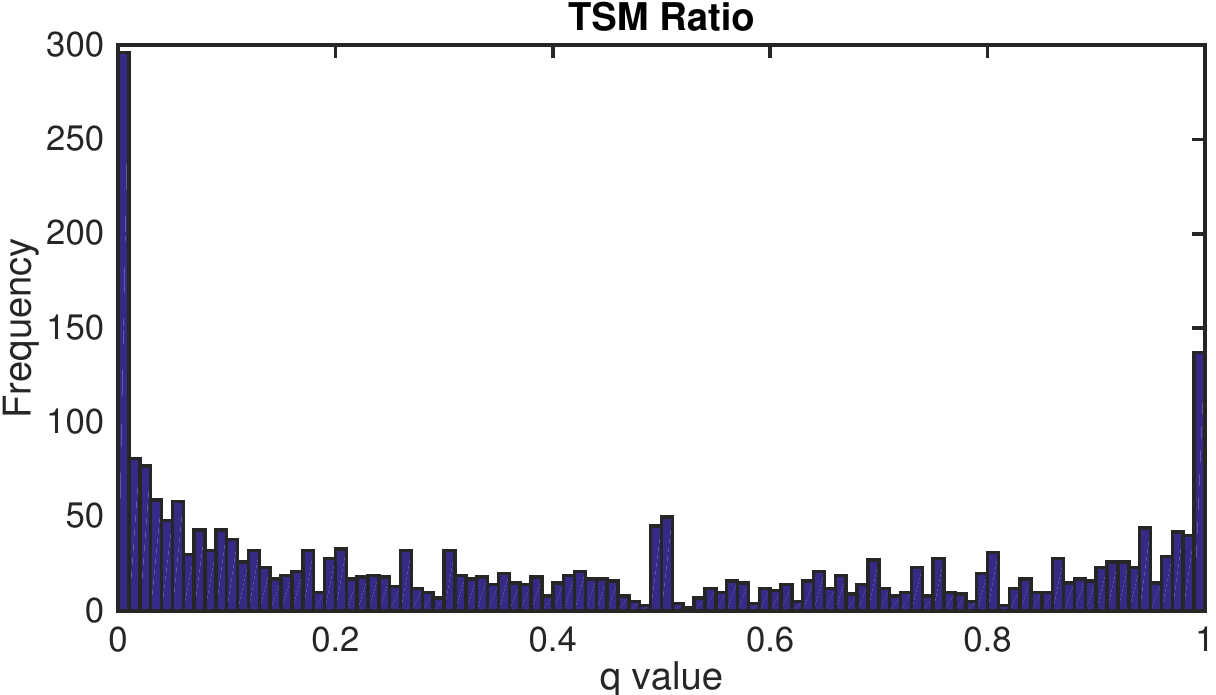}
\caption{ }
\end{subfigure}
\begin{subfigure}[b]{0.49\textwidth}
\includegraphics[width=\textwidth]{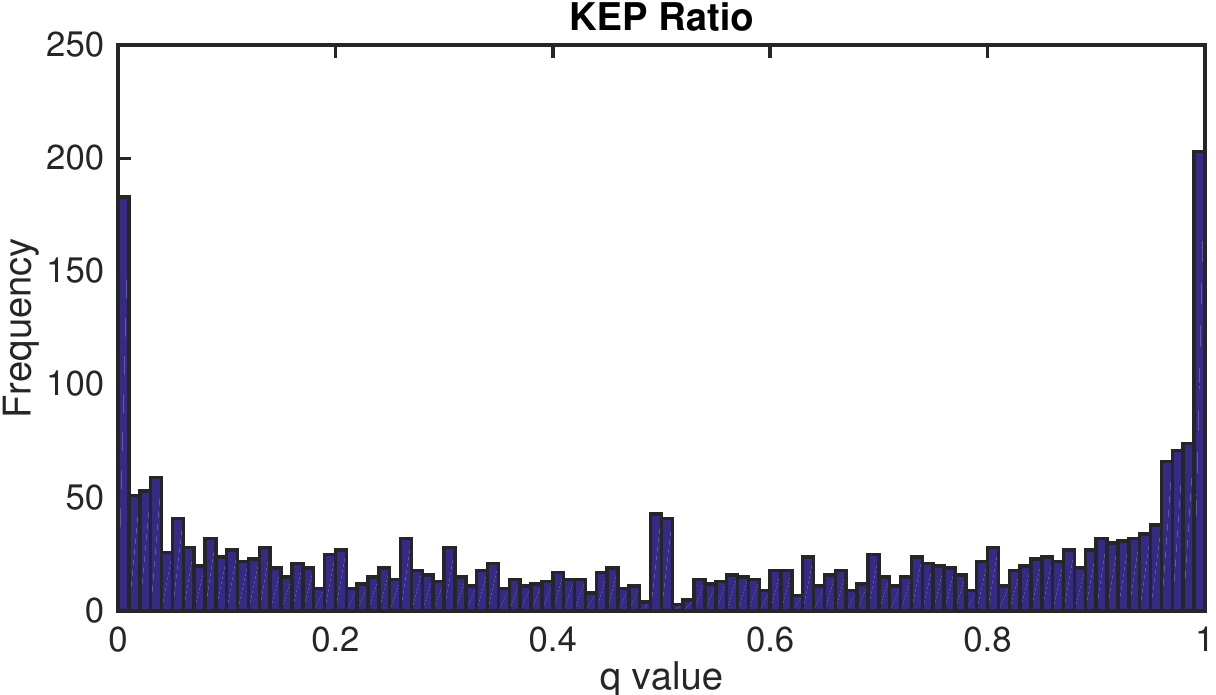}
\caption{ }
\end{subfigure}
\begin{subfigure}[b]{0.49\textwidth}
\includegraphics[width=\textwidth]{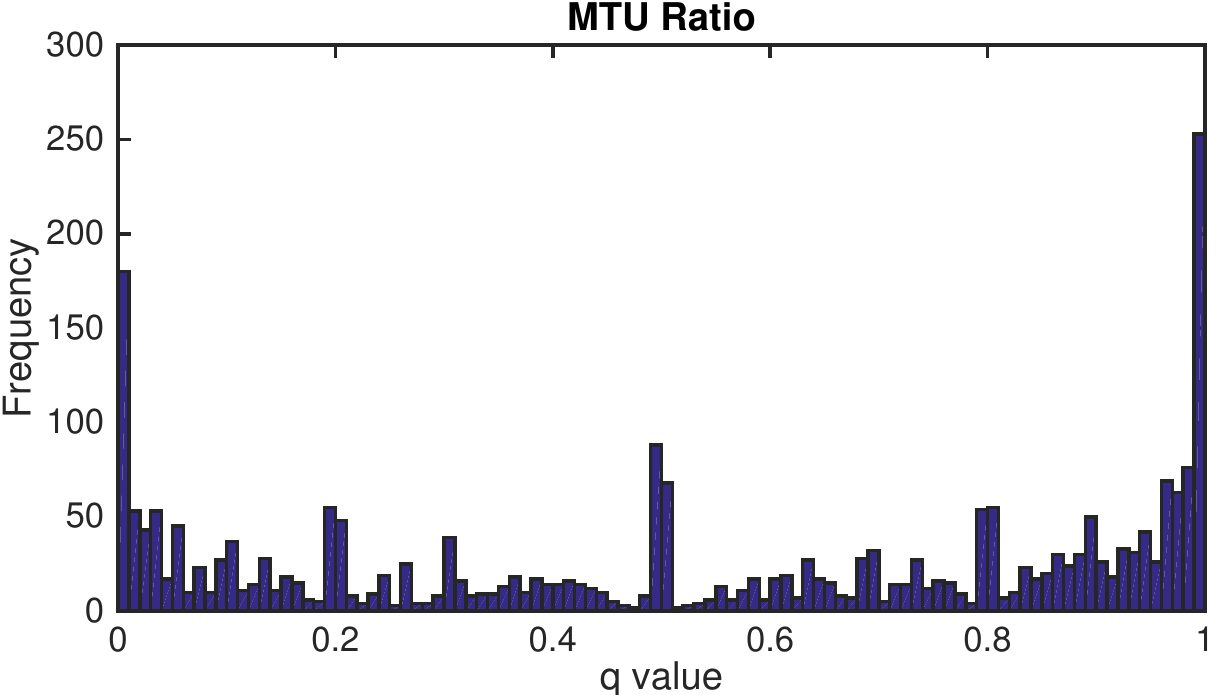}
\caption{ }
\end{subfigure}
\begin{subfigure}[b]{0.49\textwidth}
\includegraphics[width=\textwidth]{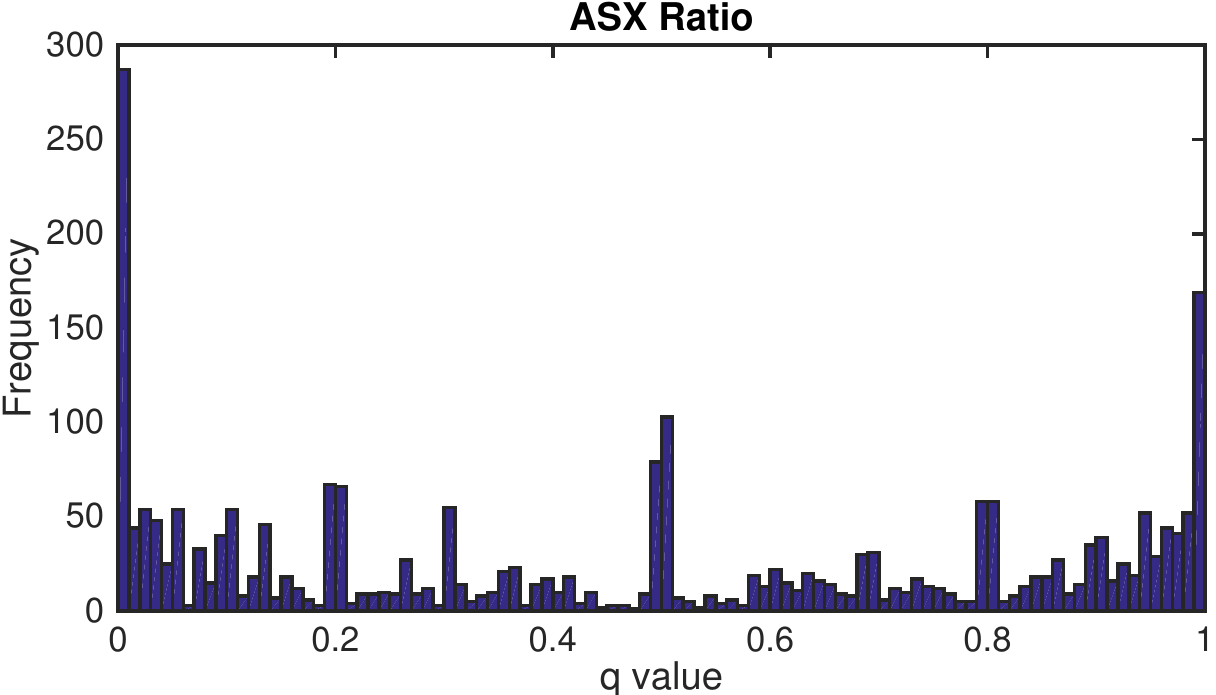}
\caption{ }
\end{subfigure}
\caption{\small{Histograms of squared returns ratio $q$ (see \eqref{eq:q})  based on empirical returns from 6/15/2004 to 6/13/2014 for ADR $\in$ $\{$(a) SHG, (b) LFC, (c) MTU, (d) ASX, (e) KEP, (f) TSM$\}$. For each histogram, the x-axis represents squared returns ratio, q, and the y-axis represents corresponding frequencies. When all the previously chosen 15 ADRs were plotted, three major shapes were observed: U-shape (6 among 15 ADRs), W-shape (7 among 15 ADRs), and two-peak-shape (MTU and ASX). Two ADRs of each case are illustrated.}}
\label{fig:Ratio}
\end{figure}

 \newpage
\begin{figure}[H]
\centering
\begin{subfigure}[H]{0.49\textwidth}
\includegraphics[height=7.7cm,keepaspectratio=true]{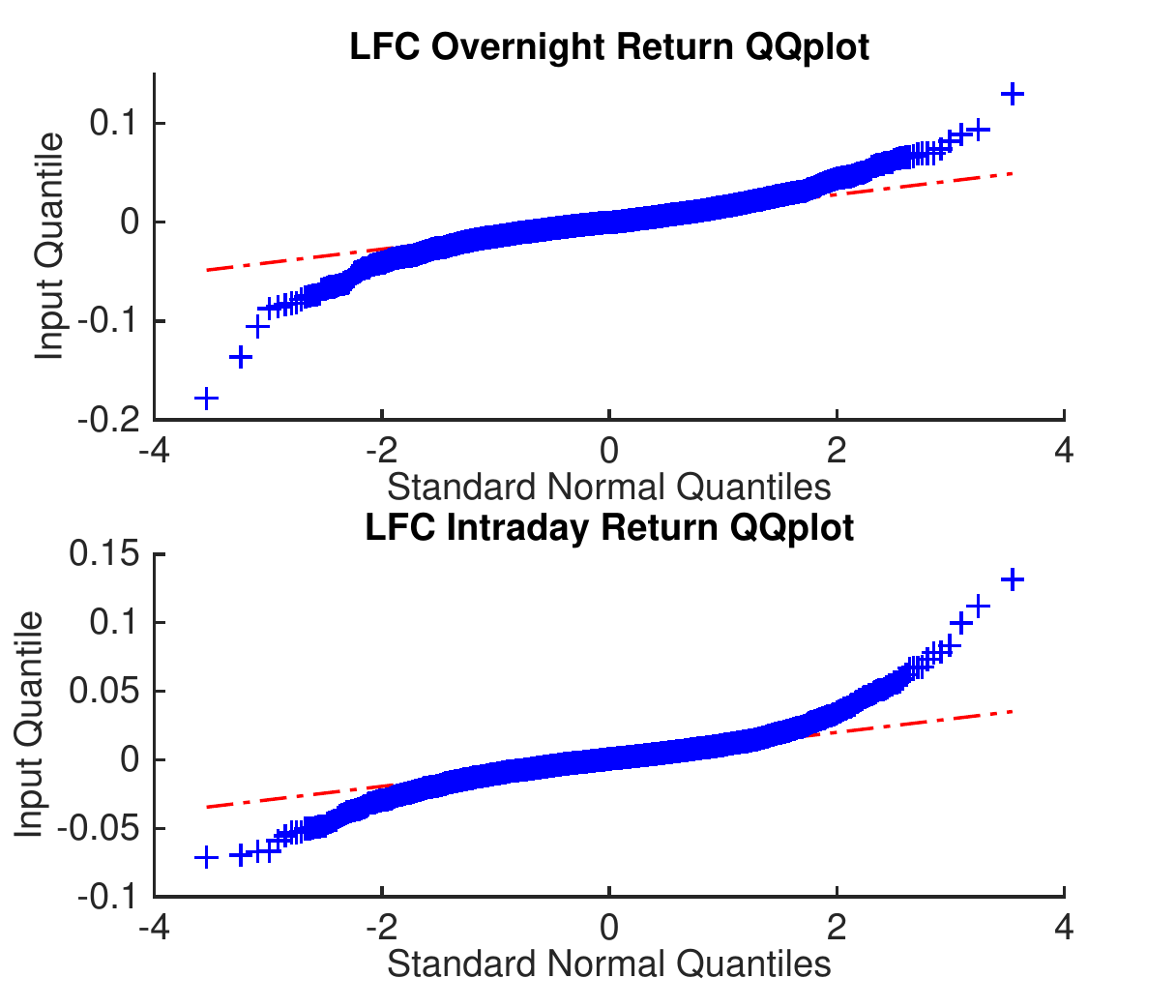}
\caption{ }
\end{subfigure}
\begin{subfigure}[H]{0.49\textwidth}
\includegraphics[height=7.7cm,keepaspectratio=true]{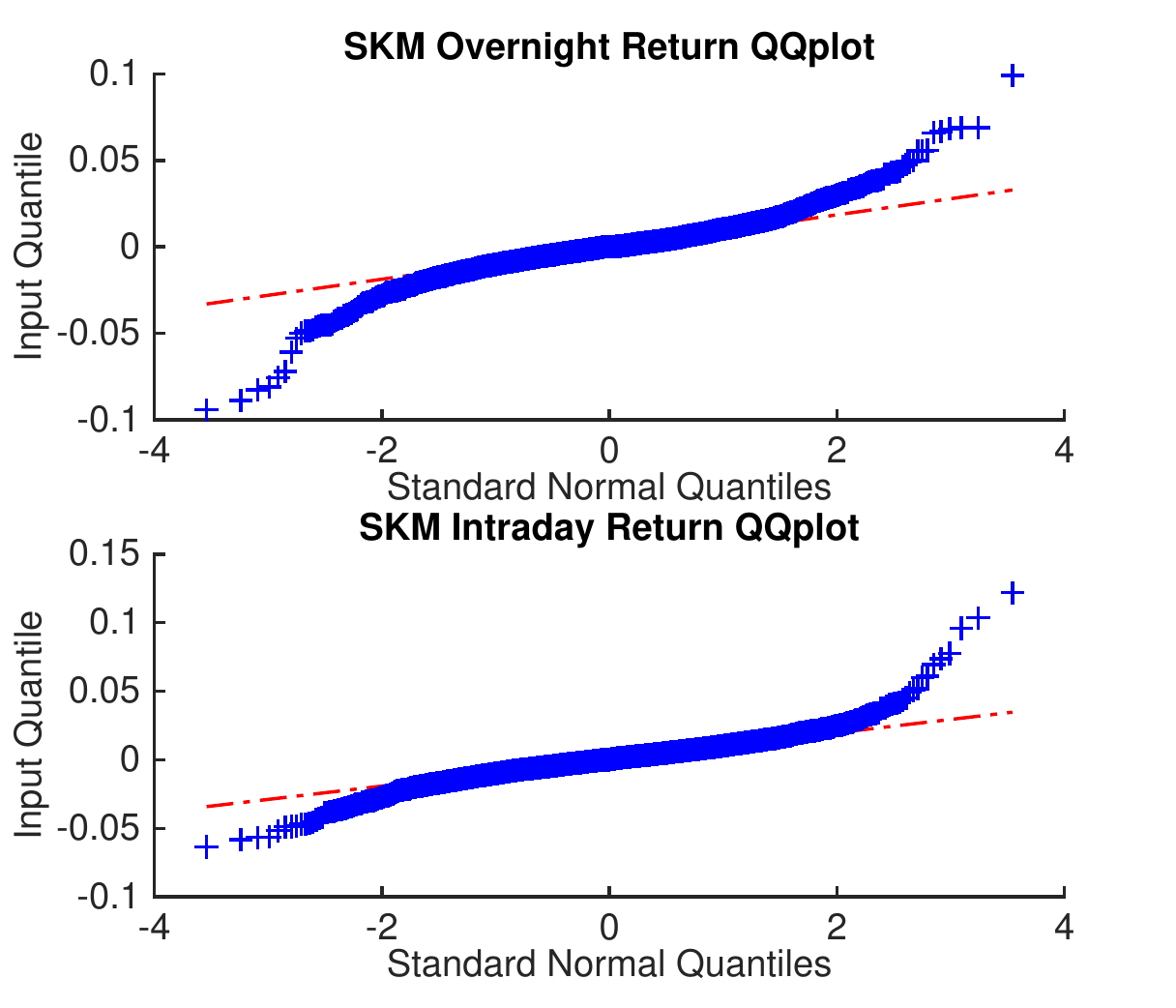}
\caption{ }
\end{subfigure}
\end{figure}%
\begin{figure}[H] 
\ContinuedFloat
\centering
\begin{subfigure}[H]{0.49\textwidth}
\includegraphics[height=7.7cm,keepaspectratio=true]{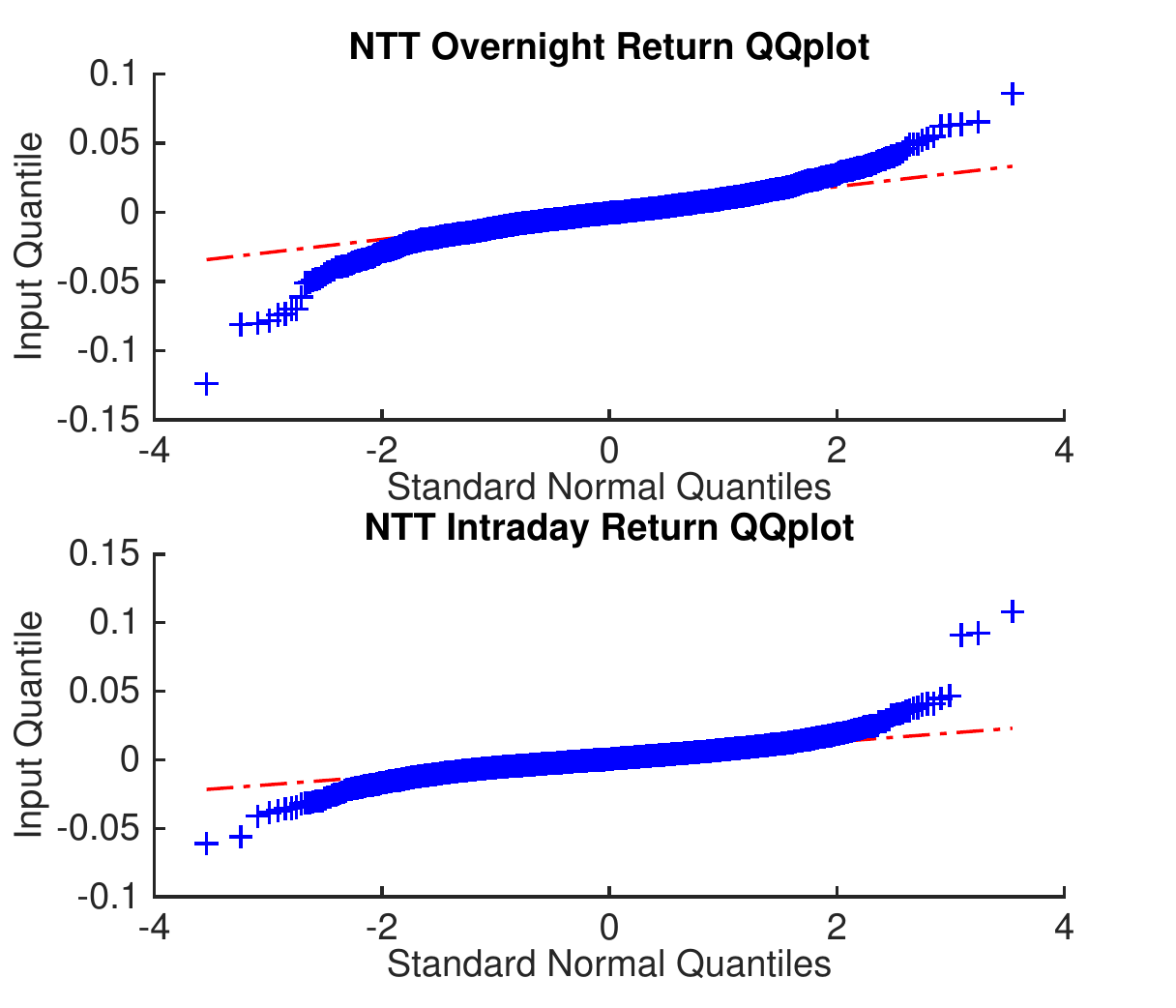}
\caption{ }
\end{subfigure}
\begin{subfigure}[H]{0.49\textwidth}
\includegraphics[height=7.7cm,keepaspectratio=true]{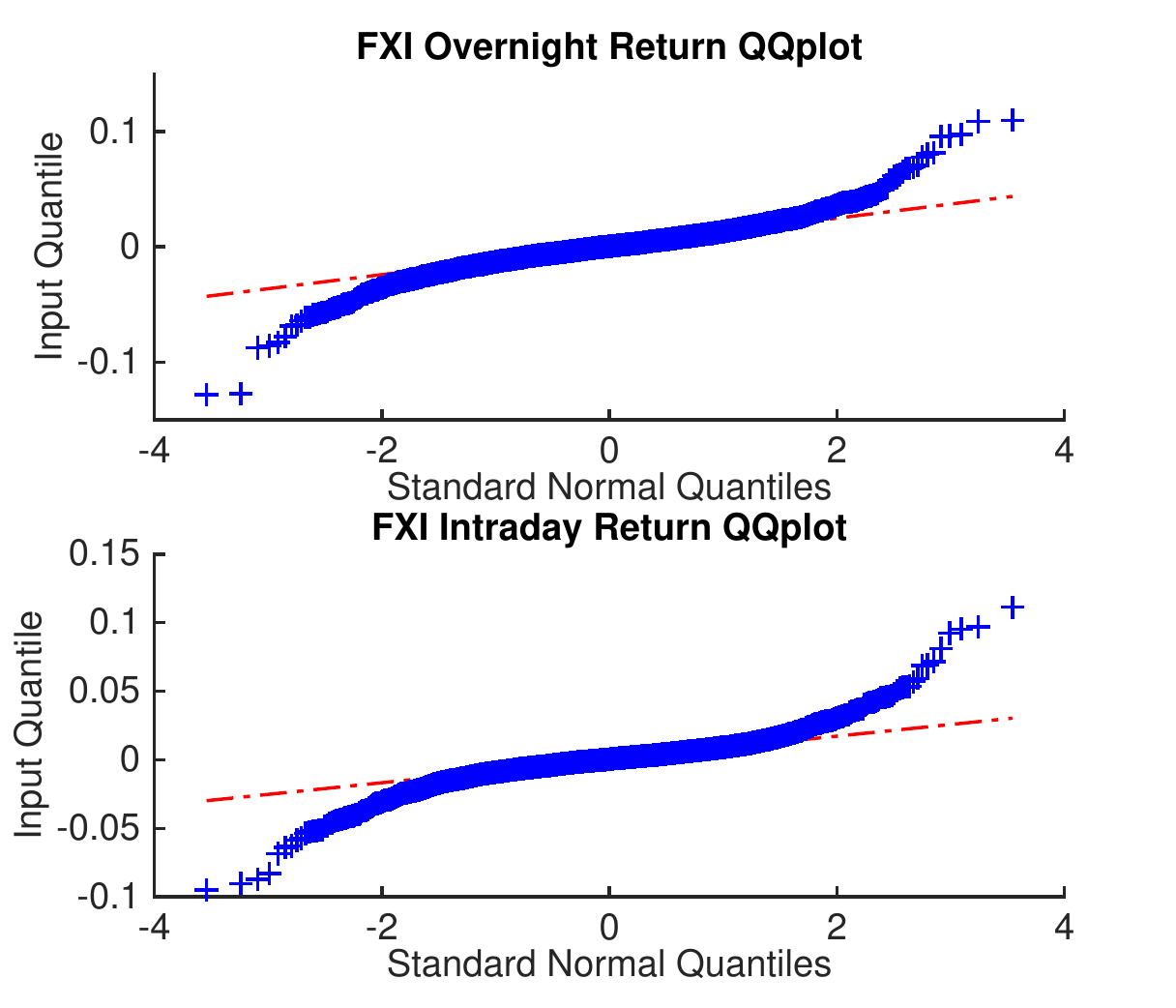}
\caption{ }
\end{subfigure}
\caption{\small{Quantile-Quantile plots of daily overnight and intraday returns from 6/15/2004 to 6/13/2014 for ADR $\in$ $\{$(a) LFC, (b) SKM, (c) NTT, (d) FXI$\}$. For each graph, the x-axis represents theoretical quantiles of a standard Normal distribution and the y-axis represents quantiles of the true distribution of the data. The closer the   crosses is to  the straight   reference line, the more normal is the empirical  distribution. }}
\label{fig:QQplot}
\end{figure}

\subsection{Testing for Normality}\label{sec2.2}
Next, we investigate whether  ADR returns follow the normal distribution. Specifically, we conduct the test  for overnight and intraday returns separately, and  the quantile-quantile plots are drawn in Figure \ref{fig:QQplot}. These plots   illustrate that the ADR returns visibly deviate from the normal distribution. Rather, their distributions appear to be heavy-tailed on both sides for all four ADRs tested: the left-tail is below and the right-tail is above the straight  reference line. This observation suggests that each ADR has extreme price fluctuations (both positively and negatively) more frequently than it would if it were normally distributed. Notice that for the intraday returns of LFC, SKM, and NTT, the degree of such deviation is more extreme on the right-side, indicating the stronger presence of positive outliers. On the other hand, we see more negative outliers (i.e. left-tails being fatter than right-tails) for the overnight returns of LFC, SKM, and NTT. For comparison, we also examine an ETF, namely FXI (iShares China Large-Cap ETF), which tends to show almost equally heavy tails on both sides. Likewise, other statistical tests may be used to further investigate their tail behaviors as well as the return distributions.

\section{ADRs and the S\&P500}\label{sec3}
A clear conclusion drawn in Section \ref{sec2} is that a large proportion of most ADR price changes occurs after the U.S. market hours (i.e. overnight). These price changes are 
less correlated with (and thus are relatively arbitrary) the U.S. market since such fluctuations tend to stem from foreign factors, such as oversees political and financial events in the originating home markets. Therefore, at least in this study, their analysis is not our main objective.

However, price fluctuations during market hours (i.e. intraday) are relatively small, yet not negligible, and much less arbitrary. Therefore, recognizing their factors and making an effective prediction of their behaviors are crucial. Earlier literature has found that ADRs follow the U.S. market during market hours. There are various indexes and ETFs that provide plausible representations of the U.S. market returns. Among them, we choose SPDR S\&P 500 ETF (SPY) as our benchmark. Since our final objective is to establish an optimal pairs-trading strategy, the focus in this section will be on the return differences between ADR and SPY (i.e. ADR-SPY), rather than the individual returns. 

\subsection{Overnight, Intraday, \& Daily Correlations}\label{sec3.1}
The most intuitive method of analyzing an $\{$ADR, SPY$\}$ pair is to evaluate the difference and correlation between the returns of ADR and SPY. From here, we denote the return spread by
\begin{equation}
X_{\alpha}(i)=R^{ADR}_{\alpha}(i)-R^{SPY}_{\alpha}(i) \qquad  \alpha \in \{ON, ID, DD\}.
\label{eq:returnspread}
\end{equation}
We expect ADR returns to have higher correlations with SPY during the U.S. market hours than they do overnight. This is due to the fact that SPY is an American ETF, being most actively traded during day hours. Hence, the U.S. market intraday returns are supposed to be the most dominant force of ADR intraday returns; whereas ADR overnight returns are most significantly affected by home market sentiments rather than the U.S. market. Therefore, we investigate the correlation between ADR and SPY returns for different time components. Then we examine the degree to which such return spreads fluctuate, further leading to our analysis of their mean-reverting stochastic behavior in Section \ref{sec3.2}. The former results are summarized in Table \ref{tab:ADRSPYstat}.

\begin{table}[H]
  \centering
  \begin{small}
  \begin{adjustbox}{width=\textwidth}
    \begin{tabular}{ll | ccc|  ccc | ccc}
    \toprule
          &       & \multicolumn{3}{c|}{\textbf{Overnight}} & \multicolumn{3}{c|}{\textbf{Intraday}} & \multicolumn{3}{c}{\textbf{Daily}} \\
                             &       & \multicolumn{3}{c|}{\textbf{$X_{ON}$}} & \multicolumn{3}{c|}{\textbf{$X_{ID}$}} & \multicolumn{3}{c}{\textbf{$X_{DD}$}}\\
          \midrule
Country & ADR & \textbf{Corr$(\%)$} & \textbf{$\mu(10^{-2}\%)$} & \textbf{$\sigma(\%)$} & \textbf{Corr$(\%)$} & \textbf{$\mu(10^{-2}\%)$} & \textbf{$\sigma(\%)$} & \textbf{Corr$(\%)$} & \textbf{$\mu(10^{-2}\%)$} & \textbf{$\sigma(\%)$} \\
    \toprule   
    \multirow{3}{*}{\textbf{CHINA}}  & \textbf{CHL}  & 53.91     & 1.71               & 1.35                  & 73.69                                                           & 2.30               & 0.87                  & 63.57                                                           & 4.20               & 1.68                  \\
                                 & \textbf{LFC}  & 55.86     & -0.25              & 1.69                  & 73.15                                                           & 6.61               & 1.02                  & 63.06                                                           & 6.61               & 2.05                  \\
                                 & \textbf{PTR}  & 60.26     & 1.08               & 1.40                  & 74.78                                                           & 2.44               & 0.96                  & 68.96                                                           & 3.78               & 1.76                  \\
\midrule
\multirow{3}{*}{\textbf{INDIA}}  & \textbf{HDB}  & 64.26     & 5.93               & 1.40                  & 60.35                                                           & 3.30               & 1.74                  & 70.39                                                           & 9.45               & 2.18                  \\
                                 & \textbf{INFY} & 52.44     & -1.34              & 1.36                  & 61.06                                                           & 4.82               & 1.34                  & 64.28                                                           & 3.63               & 1.87                  \\
                                 & \textbf{WIT}  & 48.94     & 2.61               & 1.49                  & 53.82                                                           & 2.48               & 2.01                  & 67.85                                                           & 4.49               & 2.09                  \\
\midrule
\multirow{3}{*}{\textbf{JAPAN}}  & \textbf{MTU}  & 46.53     & -12.02             & 1.64                  & 62.24                                                           & 10.67              & 1.22                  & 59.59                                                           & -1.72              & 1.84                  \\
                                 & \textbf{NTT}  & 25.09     & -8.05              & 1.36                  & 56.79                                                           & 7.24               & 0.90                  & 42.66                                                           & -0.91              & 1.56                  \\
                                 & \textbf{TM}   & 48.20     & 1.78               & 1.18                  & 76.12                                                           & -1.71              & 0.71                  & 66.90                                                           & 0.07               & 1.29                  \\
\midrule
\multirow{3}{*}{\textbf{KOREA}}  & \textbf{KEP}  & 49.50     & 2.00               & 1.57                  & 66.50                                                           & 0.90               & 1.20                  & 65.77                                                           & 3.08               & 1.93                  \\
                                 & \textbf{SHG}  & 47.26     & -0.45              & 2.00                  & 70.44                                                           & 6.04               & 1.25                  & 63.01                                                           & 5.98               & 2.40                  \\
                                 & \textbf{SKM}  & 42.52     & -2.28              & 1.24                  & 49.58                                                           & 1.84               & 1.18                  & 53.21                                                           & -0.68              & 1.53                  \\
\midrule
\multirow{3}{*}{\textbf{TAIWAN}} & \textbf{ASX}  & 35.88     & 3.53               & 1.85                  & 43.90                                                           & -0.31              & 1.73                  & 48.29                                                           & 3.13               & 2.40                  \\
                                 & \textbf{CHT}  & 32.58     & -7.78              & 1.31                  & 50.25                                                           & 8.90               & 1.15                  & 51.96                                                           & 1.03               & 1.58                  \\
                                 & \textbf{TSM}  & 56.01     & 5.45               & 1.06                  & 62.01                                                           & -2.02              & 1.37                  & 64.41                                                           & 3.37               & 1.66                  \\
 \bottomrule
 \end{tabular}
  \end{adjustbox}
  \caption{\small{Statistics of ADR and SPY returns from June 15, 2004 to June 13, 2014. First column of each time component reports the correlation coefficients between ADR and SPY returns. Second column reports the mean return spreads (ADR-SPY). Third column reports the standard deviation of the return spreads.}}
  \label{tab:ADRSPYstat}
 \end{small}
\end{table}

\newpage
Most of the ADRs considered herein  have positive intraday mean return spreads ($X_{ID}$) with high SPY correlations. Not only does this confirm our initial hypothesis that ADR prices generally move in accordance with SPY during the U.S. market hours, but it also indicates that ADRs tend to outperform SPY during these hours. Even in the cases of some exceptions, such as TM and ASX, the magnitudes of the negative mean spreads are extremely small, close to zero. On the contrary, the mean return spreads becomes arbitrarily positive and negative overnight. SPY correlations become much smaller (mostly below 50\% and all of them below their daily SPY correlations ) overnight, which points to the weaker ties between the U.S. market sentiments and ADR returns after the U.S. market hours. 

All ADRs but 3 among 15 show high correlation, greater than $40\%$, with the U.S. market especially during intraday. In particular, all 3 ADRs from China have U.S. correlation over $50\%$. China's PTR (Petro China), for example, has $60.26\%$ overnight and $74.78\%$ intraday U.S. correlations, marking itself as the most correlated among the listed ADRs. This is attributable to their large market capitalization and size of their business. Moreover, intraday U.S. correlation is also high for Japan's TM (Toyota Motors) and Korea's SHG (Shinhan Financial Group). In the case of TM, this can be explained by Toyota's international business in the United States; for SHG, consider its business sector, financial services, which can be highly influenced by the U.S. market sentiments. Similarly, Japan's MTU (Mitsubishi UFJ Financial Group) has a high U.S. intraday correlation. As such, business sector tends to have significant influence on the correlation (especially intraday) between the corresponding ADR and the U.S. market. There are some exceptions, however. For instance, Taiwan's ASX (Advanced Semiconductor Engineering) and TSM (Taiwan Semiconductor Manufacturing) have fairly different U.S. correlations: ASX has $35.88\%$ overnight and $43.90\%$ intraday, whereas TSM has $56.01\%$ overnight and $62.01\%$ intraday. Factors other than home country and business sector might have played a role here, which calls for future investigations. Furthermore, all but 3 ADRs have standard deviations of return spreads (ADR-SPY) higher during overnight periods than during intraday, again implying the tight relationship between ADRs and the U.S. market during U.S. market hours.

\subsection{Mean-Reverting Spreads}\label{sec3.2}
In this section, we take a closer look at the time series of the ADR-SPY spreads. In particular, we examine the   stochastic behaviors of the pairs: Do they have a consistent upward/downward trend? Are they mean-reverting? Figure \ref{fig:ADR-SPYplot} illustrates the return spread time series using price data during 10/1/2013--2/24/2014  (100 days).

The return spreads in Figure \ref{fig:ADR-SPYplot} display rather volatile movements. In particular,  every abrupt decline is often followed by an upward spike, which is a typical behavior of a mean-reverting process. We model the return spreads by  an Ornstein-Uhlenbeck process, described by the stochastic differential equation
\begin{equation}
dX(t) = \mu(\theta-X(t))dt + \sigma dW(t)\,,
\label{eq:OU}
\end{equation}
where $X(t)\in\mathbb{R}$ is the  return spread at time $t$, $\mu>0$ is the  rate of mean reversion, $\theta\in\mathbb{R}$ is the  long-term mean, and $\sigma>0$ is the volatility parameter. The process is driven by a standard Brownian motion, denoted by  $W(t)$. We consider the daily returns as realization of this mean-reverting process. We can then express the conditional probability of $X(t_i)$, given previous data $x(i-1)$ at time $t(i-1)$ with time step $\Delta t=t(i)-t(i-1)=1$, as
\begin{align}
f^{OU}(x(i)|x(i-1);\theta,\mu,\sigma) = \frac{1}{\sqrt{2\pi\tilde{\sigma}^2}}&\exp\left(-\frac{(x(i)-x(i-1)e^{-\mu \Delta t} - \theta(1-e^{-\mu \Delta t}))^2 }{2\tilde{\sigma}^2} \right) \,,
\end{align}
\text{where }
\begin{equation}
\tilde{\sigma}^2 = \sigma^2\frac{1-e^{-2\mu \Delta t}}{2\mu}\,.
 \end{equation}
An observation of the sequence  $(x(1),x(2),\dots,x(N))$ where $N=2516$ allows us to compute, and thereby maximize the average log-likelihood defined  by
\begin{equation}
\begin{split}
\ell(\theta,\mu,\sigma|x(1),\dots, x(N)) &:= \frac{1}{n}\sum_{i=1}^{n}\ln f^{OU}\left(x(i)|x(i-1);\theta,\mu,\sigma \right)\notag\\
&=-\frac{1}{2}\ln(2\pi)-\ln(\tilde{\sigma}) \notag \\
&\quad -\frac{1}{2n\tilde{\sigma}^2}\sum_{i=1}^{n}[x(i)-x(i-1)e^{-\mu \Delta t} - \theta(1-e^{-\mu \Delta t} )]^2.
\end{split}
\end{equation}

Adopting the same maximum likelihood estimation (MLE) methodology used in  \cite{LeungLi2015OU}, we attain the following parameter values for the fitted OU process. The values are first estimated using empirical price data (from June 2004 to June 2014) of the chosen ADR and SPY. Then we simulate the OU process using the estimated parameters from the first round and re-estimate for parameters. For example, we report the former and the latter results of the return spread TSM-SPY in Table \ref{tab:TSMMLE} along with their maximized log-likelihood values. All values are reported for every calendar year from 2005 to 2014. Empirical  returns during  6/15/2004--6/13/2014 are used.  As shown, the estimated parameters from the empirical return spreads and from the simulated OU process are quite close, especially in the case of the intraday spreads, which indicates that they fit to the OU model relatively well. Also, based on the empirical  estimates we can see that the spreads tend to revert to the long-term mean faster during U.S. market hours than they do overnight. However, there are some discrepancies between empirical and re-simulated estimates of the mean-reverting rates, which calls for a better model to capture them. \\

\begin{table}[H]
\centering
   \begin{small}
   \centering
    \hspace*{-0.5cm}\tabcolsep=0.11cm \begin{tabular}{r | cccc | cccc | cccc}
    \toprule
         Year & \multicolumn{4}{c|}{\textbf{Overnight}} & \multicolumn{4}{|c|}{\textbf{Intraday}} & \multicolumn{4}{|c}{\textbf{Daily}} \\
    & \textbf{$\hat{\theta}$} & \textbf{$\hat{\sigma}$} & \textbf{$\hat{\mu}$} & \textbf{$\hat{\ell}$} & \textbf{$\hat{\theta}$} & \textbf{$\hat{\sigma}$} & \textbf{$\hat{\mu}$} & \textbf{$\hat{\ell}$} & \textbf{$\hat{\theta}$} & \textbf{$\hat{\sigma}$} & \textbf{$\hat{\mu}$} & \textbf{$\hat{\ell}$} \\
    \midrule
    \multicolumn{1}{l}{} & \multicolumn{12}{c}{\textbf{Empirical}} \\
    \midrule
	\textbf{2005} & 0.00081  & 0.0272  & 2.9480 & 3.0733   & -0.00004 & 0.0338  & 3.7592 & 3.0487   & 0.00075  & 0.0488  & 4.6075 & 2.7610   \\
	\textbf{2006} & 0.00213  & 0.0254  & 2.4819 & 3.0597   & -0.00206 & 0.0362  & 2.7291 & 2.8677   & 0.00005  & 0.0473  & 3.5031 & 2.6854   \\
	\textbf{2007} & 0.00062  & 0.0315  & 3.9448 & 3.1380   & -0.00109 & 0.0417  & 5.3822 & 2.9461   & -0.00047 & 0.0362  & 2.0693 & 2.7899   \\
	\textbf{2008} & -0.00163 & 0.0357  & 2.9581 & 2.8051   & 0.00318  & 0.0555  & 3.4375 & 2.4365   & 0.00152  & 0.0588  & 2.0440 & 2.3012   \\
	\textbf{2009} & 0.00014  & 0.0230  & 1.5803 & 2.9508   & 0.00079  & 0.0594  & 4.7427 & 2.5301   & 0.00088  & 0.0569  & 3.6794 & 2.4466   \\
	\textbf{2010} & 0.00112  & 0.0164  & 2.8408 & 3.6695   & -0.00115 & 0.0270  & 3.5113 & 3.1693   & -0.00002 & 0.0352  & 3.8312 & 3.0136   \\
	\textbf{2011} & 0.00002  & 0.0173  & 2.6537 & 3.4745   & 0.00023  & 0.0288  & 4.0082 & 3.1706   & 0.00026  & 0.0339  & 2.4709 & 2.9023   \\
	\textbf{2012} & 0.00053  & 0.0211  & 2.7700 & 3.4112   & 0.00030  & 0.0334  & 4.4183 & 3.1242   & 0.00083  & 0.0333  & 2.4038 & 2.9118   \\
	\textbf{2013} & 0.00082  & 0.0180  & 1.8182 & 3.2562   & -0.00186 & 0.0272  & 5.0457 & 3.3419   & -0.00103 & 0.0306  & 2.5400 & 2.8842   \\
	\textbf{2014} & -0.00020 & 0.0183  & 2.6232 & 3.5381   & 0.00168  & 0.0219  & 2.6881 & 3.3626   & 0.00148  & 0.0281  & 1.8906 & 3.0233   \\
    \midrule
     \multicolumn{1}{l}{} & \multicolumn{12}{c}{\textbf{Simulated}} \\
    \midrule
	\textbf{2005} & 0.00072  & 0.0369  & 4.6138 & 2.9923   & 0.00059  & 0.0288  & 2.8673 & 3.0033   & 0.00117  & 0.0480  & 4.7032 & 2.7388   \\
	\textbf{2006} & 0.00343  & 0.0283  & 2.6396 & 3.1027   & -0.00209 & 0.0384  & 3.4938 & 2.8139   & -0.00067 & 0.0457  & 3.2263 & 2.5997   \\
	\textbf{2007} & 0.00133  & 0.0265  & 1.9512 & 3.0877   & -0.00123 & 0.0382  & 4.1153 & 2.9003   & -0.00065 & 0.0401  & 2.6532 & 2.7541   \\
	\textbf{2008} & -0.00292 & 0.0379  & 2.1535 & 2.7537   & 0.00221  & 0.0556  & 3.1385 & 2.4846   & -0.00216 & 0.0665  & 3.7474 & 2.3578   \\
	\textbf{2009} & 0.00038  & 0.0300  & 2.4357 & 2.8832   & 0.00019  & 0.0519  & 2.8585 & 2.5224   & -0.00149 & 0.0605  & 4.2882 & 2.4601   \\
	\textbf{2010} & 0.00206  & 0.0170  & 3.7433 & 3.7322   & -0.00125 & 0.0412  & 7.3943 & 3.1397   & -0.00111 & 0.0355  & 3.7535 & 2.9985   \\
	\textbf{2011} & -0.00022 & 0.0224  & 4.2073 & 3.5027   & 0.00084  & 0.0277  & 3.2434 & 3.1915   & -0.00027 & 0.0316  & 2.6965 & 2.8808   \\
	\textbf{2012} & 0.00050  & 0.0249  & 4.2419 & 3.4006   & 0.00019  & 0.0323  & 4.3954 & 3.1545   & 0.00167  & 0.0434  & 5.3901 & 2.9430   \\
	\textbf{2013} & 0.00080  & 0.0230  & 2.7312 & 3.3202   & -0.00158 & 0.0253  & 3.1241 & 3.2710   & -0.00129 & 0.0355  & 3.1502 & 2.8398   \\
	\textbf{2014} & -0.00069 & 0.0202  & 4.8150 & 3.6174   & 0.00228  & 0.0184  & 1.7970 & 3.4345   & 0.00109  & 0.0214  & 2.1496 & 3.1620   \\
    \bottomrule
    \end{tabular}
     \caption{\small{Estimated OU parameters  of TSM-SPY. Parameters are first estimated from empirical data via MLE, and then we estimate again from the simulated OU process associated with the   parameters in the first round. Every row shows its corresponding year's estimates of long-term mean, volatility, rate of mean reversion, and maximum log-likelihood.}} 
     \label{tab:TSMMLE}
  \end{small}
\end{table}

\newpage 

\begin{figure}[H]
\caption{\small{Time series of return spreads ADR-SPY during 100 days between Oct 1, 2013 and Feb 24, 2014 for ADR $\in$ $\{$(a) CHL, (b) SKM, (c) TM, (d) ASX$\}$.}}
\centering
\begin{subfigure}[H]{0.45\textwidth}
\includegraphics[width=\textwidth]{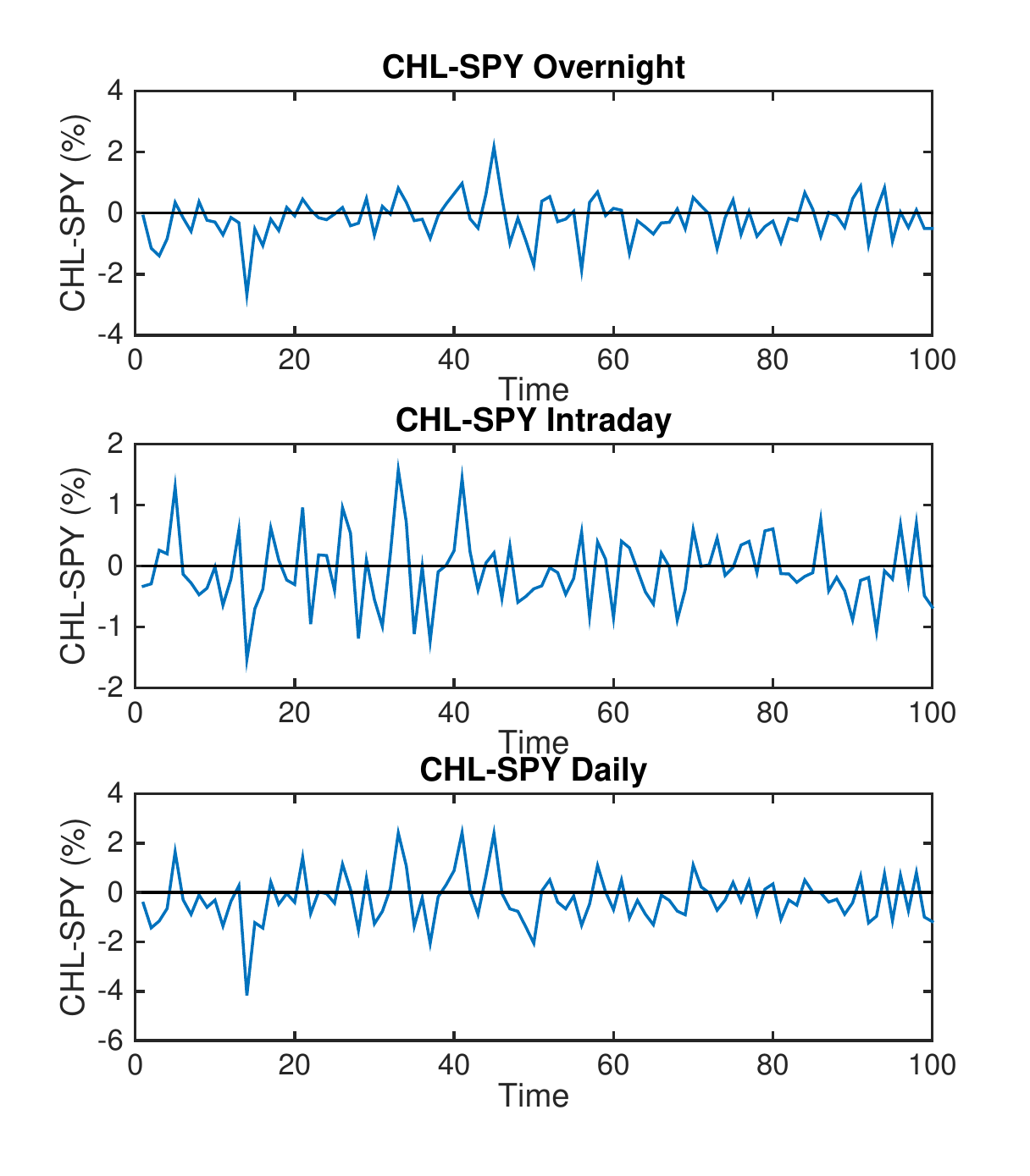}
\caption{ }
\end{subfigure}
\begin{subfigure}[H]{0.45\textwidth}
\includegraphics[width=\textwidth]{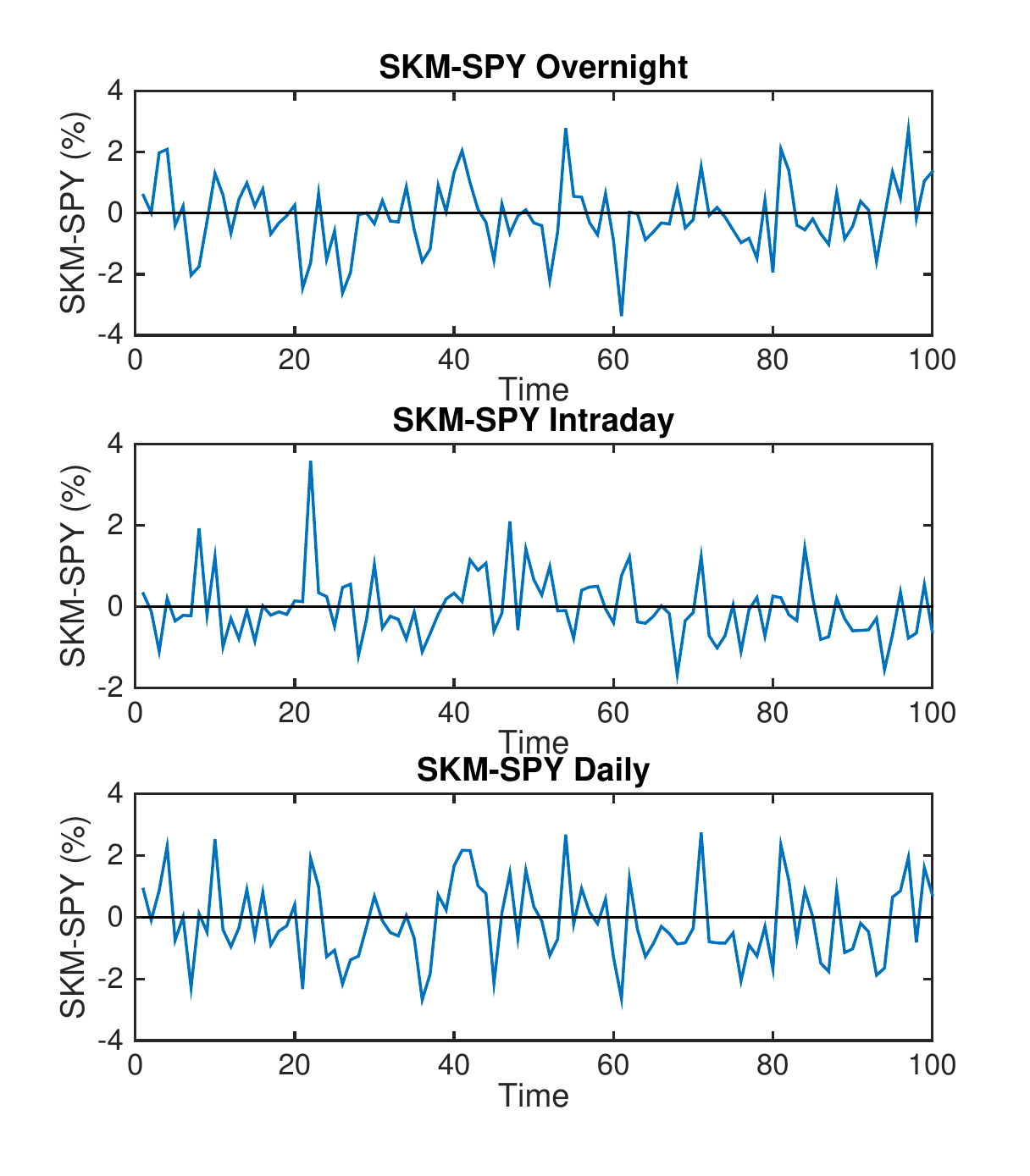}
\caption{ }
\end{subfigure}
\end{figure}
\begin{figure}[H] 
\ContinuedFloat
\centering
\begin{subfigure}[H]{0.45\textwidth}
\includegraphics[width=\textwidth]{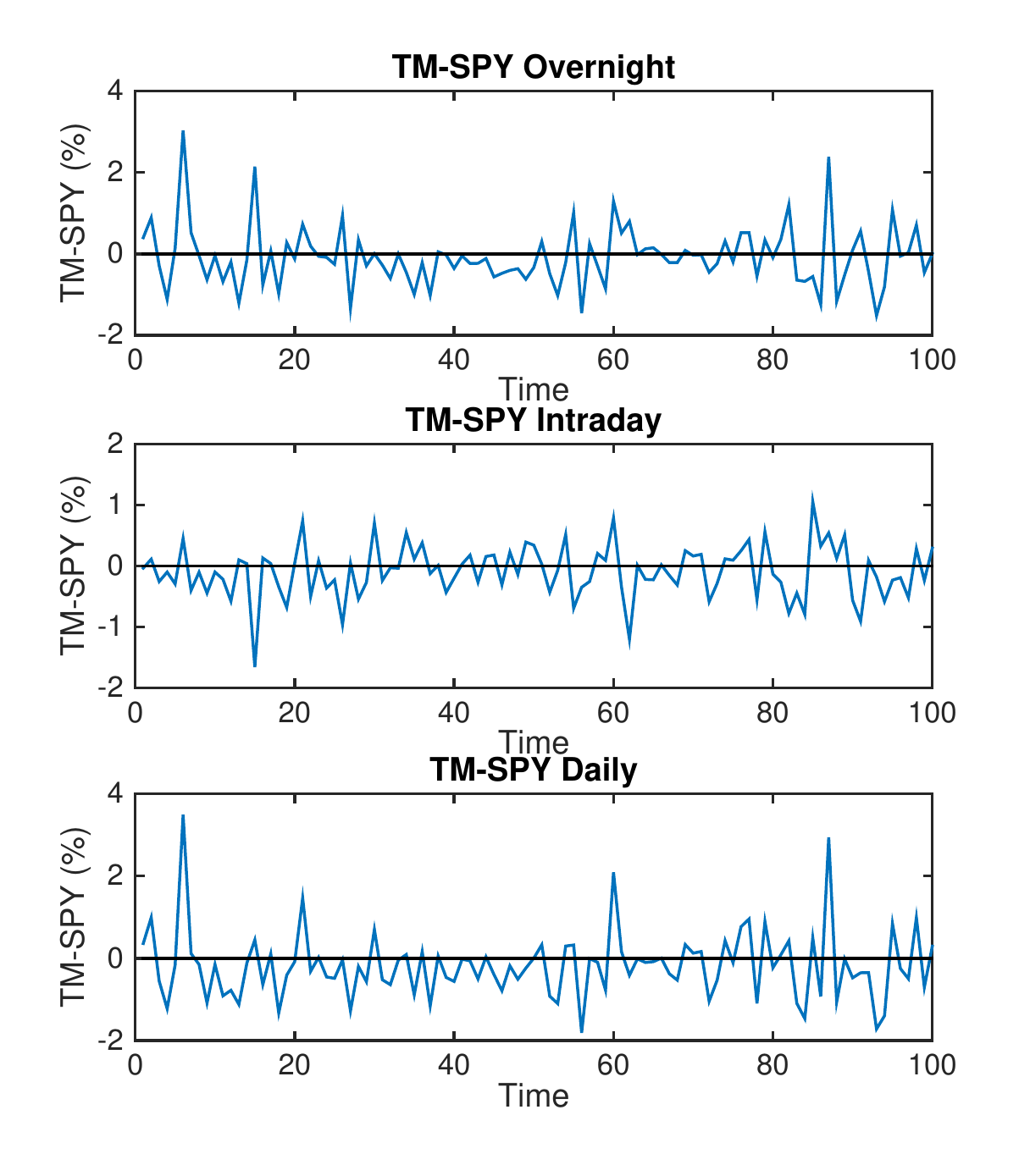}
\caption{ }
\end{subfigure}
\begin{subfigure}[H]{0.45\textwidth}
\includegraphics[width=\textwidth]{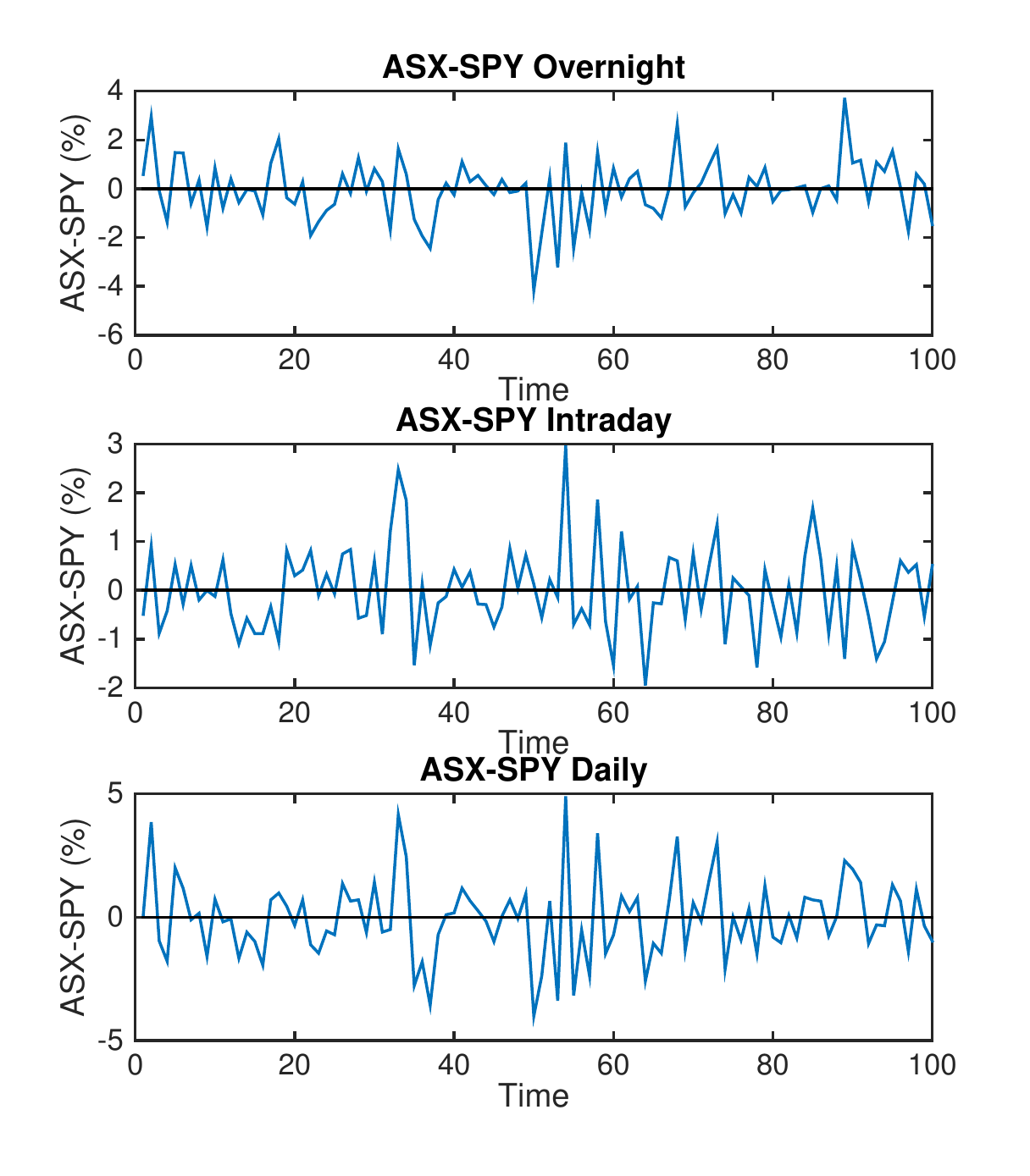}
\caption{}
\end{subfigure}
\label{fig:ADR-SPYplot}
\end{figure}


\section{ADR-SPY Pairs Trading Strategies}\label{sec4}
From Section \ref{sec3.1}, it has been identified that ADRs are likely to outperform SPY during market hours. In addition to this, now we have recognized the mean-reverting behavior of the return spread (ADR-SPY). Using these facts, we build a pairs trading strategy where we long ADR and short SPY. For an optimal result we set a specific entry level threshold and enter the given position when the return spread, ADR-SPY, is below this certain threshold value $k\%$.  It is expected that the spread ADR-SPY will soon escalate back due to its mean-reverting nature. However, the speed of this reversion is yet to be fully certain, and thus we also vary days to liquidate (i.e. exit level) along with different $k\%$ entry levels. Multiple trades can overlap with each other, meaning that each trade does not necessarily have to start once the old ones are closed. We assume here that both long and short positions are worth \$100, resulting in zero net cash outflow. Therefore, we report returns not in percentages, but instead in the cash values of their payoffs. 

In this section, we examine the pairs trading strategy by varying the entry threshold from 0\% to -1.5\% with a -0.5\% step and the number of days to liquidate from 1 to 5 days. We assume that each position is entered and closed only at market close.

\subsection{Profitability of Mean Reverting Pairs Trading Strategy}\label{sec4.1}
We have previously constructed a pairs trading strategy to long ADR and short SPY (i.e. enter ADR-SPY) when the return spread is under k\%.  The  profitability of such a strategy can be estimated through backtesting with empirical data from June 15, 2004 to June 13, 2014. For a clearer conclusion, we normalize the return from each entry/exit levels using two different schemes: 1) Annual return method, where the sum of profits during the entire 10 year data span is divided by 10, and 2) Per-trade return method, where the sum of profits is divided by the number of trades:
\begin{equation}
\begin{split}
\text{Annual Return} &= \frac{1}{10}\sum_{i=1}^{n}\text{Profit}_i \\
\text{Per-Trade Return} &= \frac{1}{n}\sum_{i=1}^{n}\text{Profit}_i \\
\text{Profit}_i = \alpha_{i}(P_{i}^{ADR}-P_{i-1}^{ADR})& - \beta_{i}(P_{i}^{SPY}-P_{i-1}^{SPY}) \\
\text{s.t. } \alpha_{i}P_{i-1}^{ADR}&=\beta_{i}P_{i-1}^{SPY}=\$100
\end{split}
\label{eq:profit}
\end{equation}
where $n$ is the number of trades, $\alpha_i$ is the number of shares of ADR long position for the $i^{th}$ trade, and $\beta_i$ is the  number of shares of SPY short position for the $i^{th}$ trade. Also, as outlined earlier, each ADR is examined for 4 different $k$ values, starting from 0\% and decreasing by $\Delta k$. The value of $\Delta k$ can be uniquely assigned to each ADR such that the trading frequencies are reasonably consistent throughout different ADRs. For all ADRs, we set $\Delta k = -0.5\%$ for convenience. By doing so, ADRs are tested with $k$ values of 0\%, -0.5\%, -1\%, and -1.5\%. This results in, for instance, 1255, 854, 543, and 315 trades respectively out of 2516 possible trading days  from 6/15/2004 to 6/13/2014 in the case of CHL. In other words, for each $k$ value, we trade approximately once every two, three, five, and eight days respectively. 

In Figure \ref{fig:trading1}, we scatter plot the normalized profit,  in terms of  mean and standard deviation, associated with  every entry threshold ($k\%$) and trade duration. Recall that an investor enters long position of \$100 in ADR and short position of \$100 in SPY, whenever the return spread, ADR-SPY, below $k\%$ is observed. Then the investor exits the positions after preassigned ``trade duration" days. For $k=\{0\%, -0.5\%, -1\%, -1.5\%\}$, all ADRs but 4 (Japan's TM, MTU, NTT and Korea's SKM) among 15 generate consistently positive payoffs. Recall from \ref{tab:ADRSPYstat} that all these 4 ADRs have either negative or extremely small daily return spread ADR-SPY. Hence, lower $k$ values seem to be more appropriate. We examine MTU in Figure \ref{fig:trading1} with two different sets of $k$ values: (e),(f) with $k=\{0\%, -0.5\%, -1\%, -1.5\%\}$ and (g),(h) with $k=\{-1\%, -1.5\%, -2\%, -2.5\%\}$. As expected, with lower entry levels, MTU generates much higher and positive returns with lower trading frequencies in (g) and (h). In general, more negative entry levels generate relatively lower annual returns but higher per trade returns. This is attributable to their small number of trades as it becomes more unlikely for the return spread to decrease below the already low $k$. Likewise, lower per trade returns are usually compensated by more frequent trades and thereby higher annual returns, generated by less negative entry levels. In addition, notice that high returns come at the cost of high standard deviations. In general, the longer the trade duration is (i.e. higher exit level), the higher both mean and standard deviation of the return are. More negative entry level also contributes to high standard deviations. It is therefore crucial to understand the tradeoff between annual vs per trade returns as well as mean vs standard deviation and carefully choose the appropriate entry/exit levels according to the investor's goal.

\begin{figure}[H]
\begin{subfigure}[b]{0.5\textwidth}
\includegraphics[width=\textwidth]{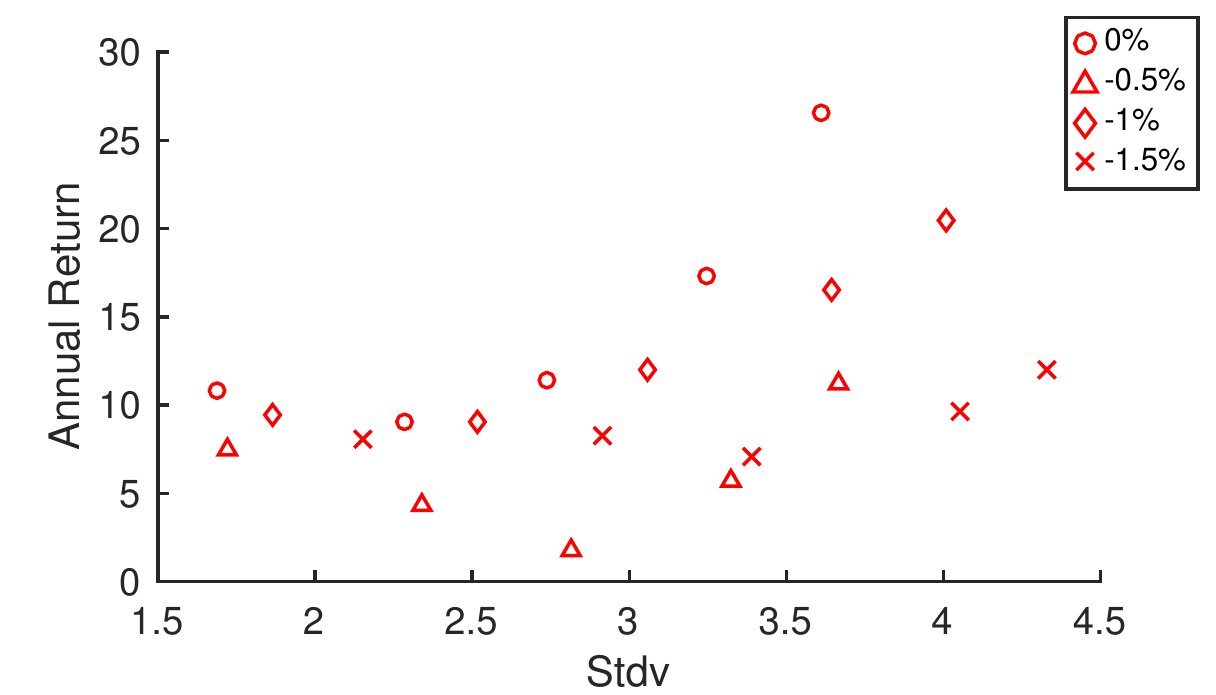}
\caption{ }
\end{subfigure}
\begin{subfigure}[b]{0.5\textwidth}
\includegraphics[width=\textwidth]{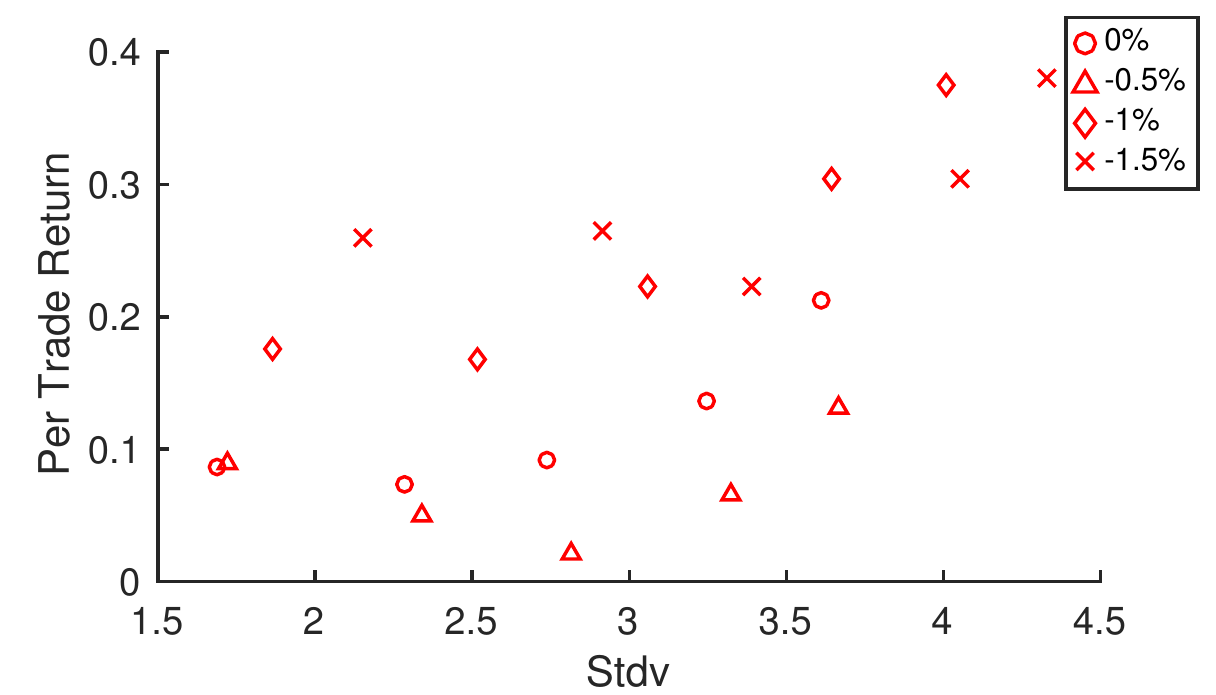}
\caption{ }
\end{subfigure}
\begin{subfigure}[b]{0.5\textwidth}
\includegraphics[width=\textwidth]{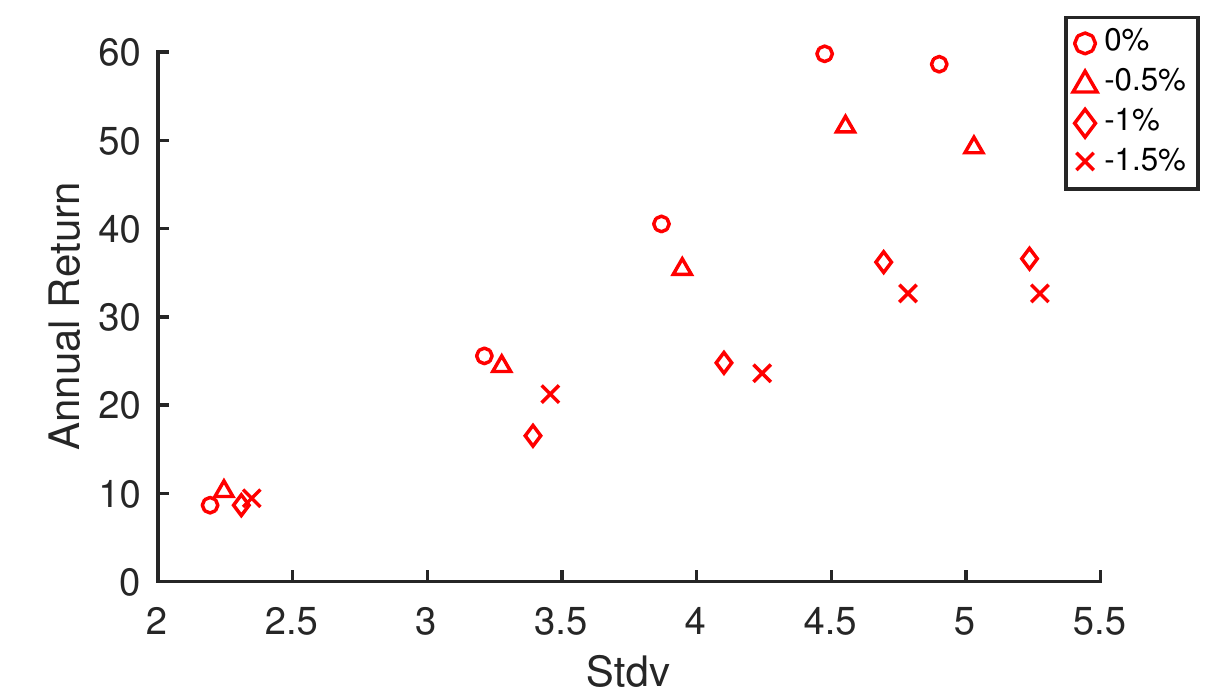}
\caption{ }
\end{subfigure}
\begin{subfigure}[b]{0.5\textwidth}
\includegraphics[width=\textwidth]{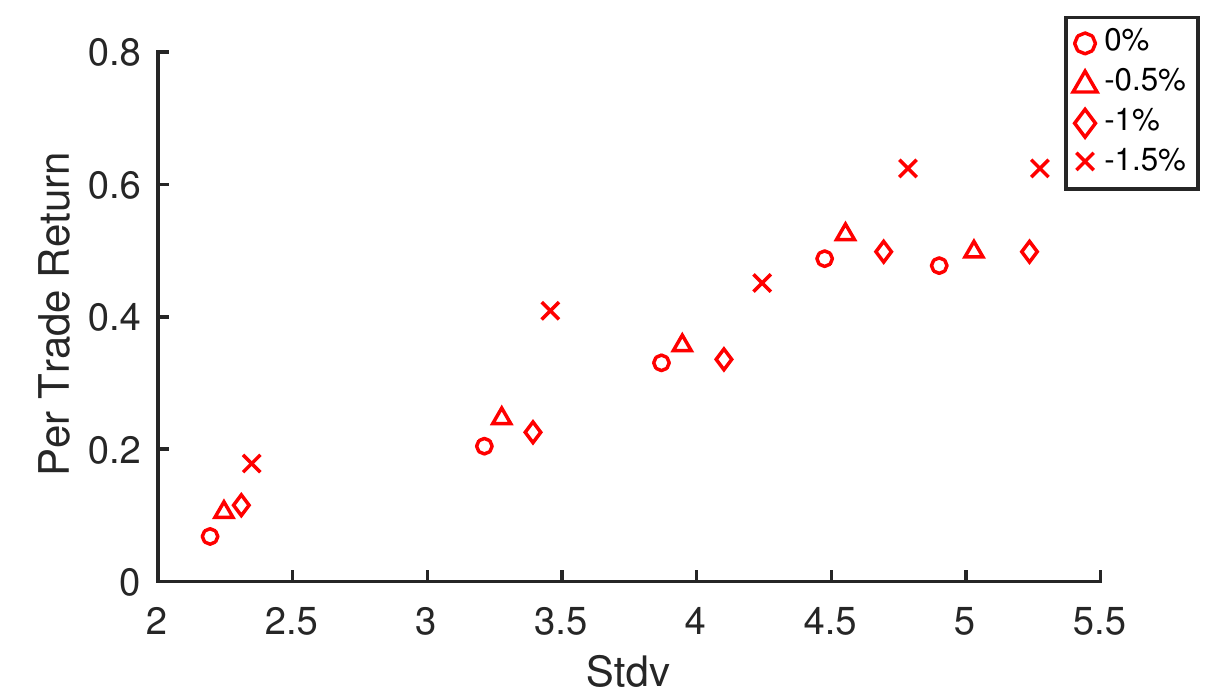}
\caption{ }
\end{subfigure}
\begin{subfigure}[b]{0.5\textwidth}
\includegraphics[width=\textwidth]{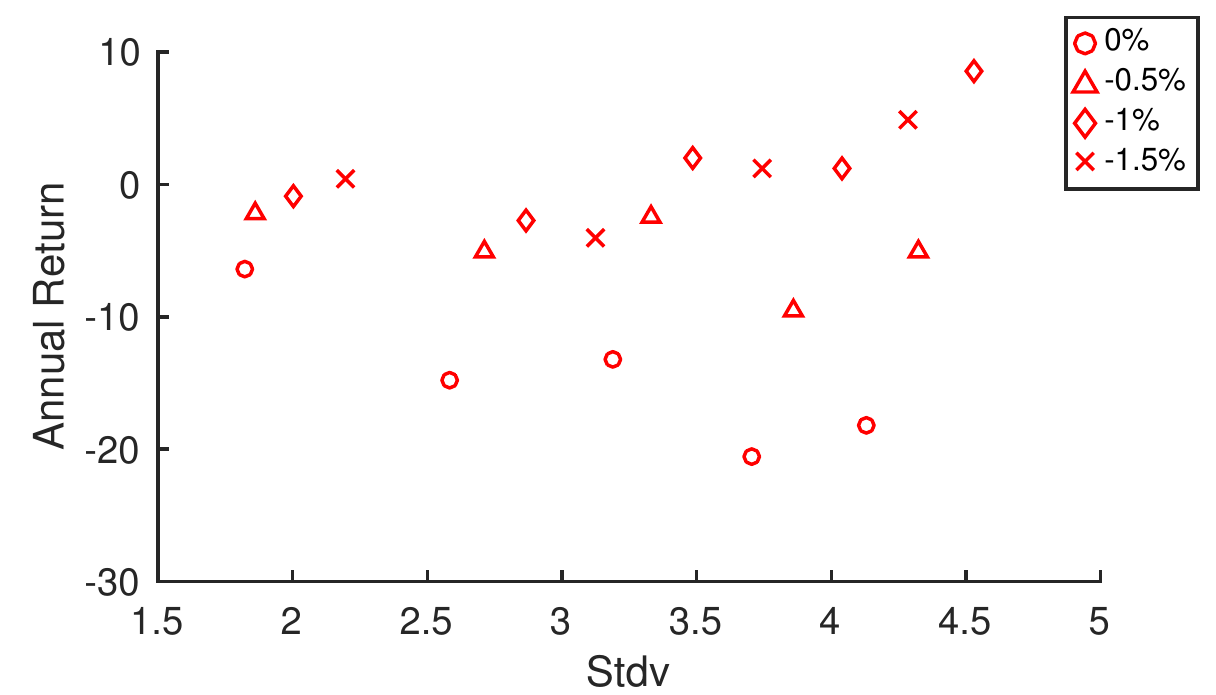}
\caption{ }
\end{subfigure}
\begin{subfigure}[b]{0.5\textwidth}
\includegraphics[width=\textwidth]{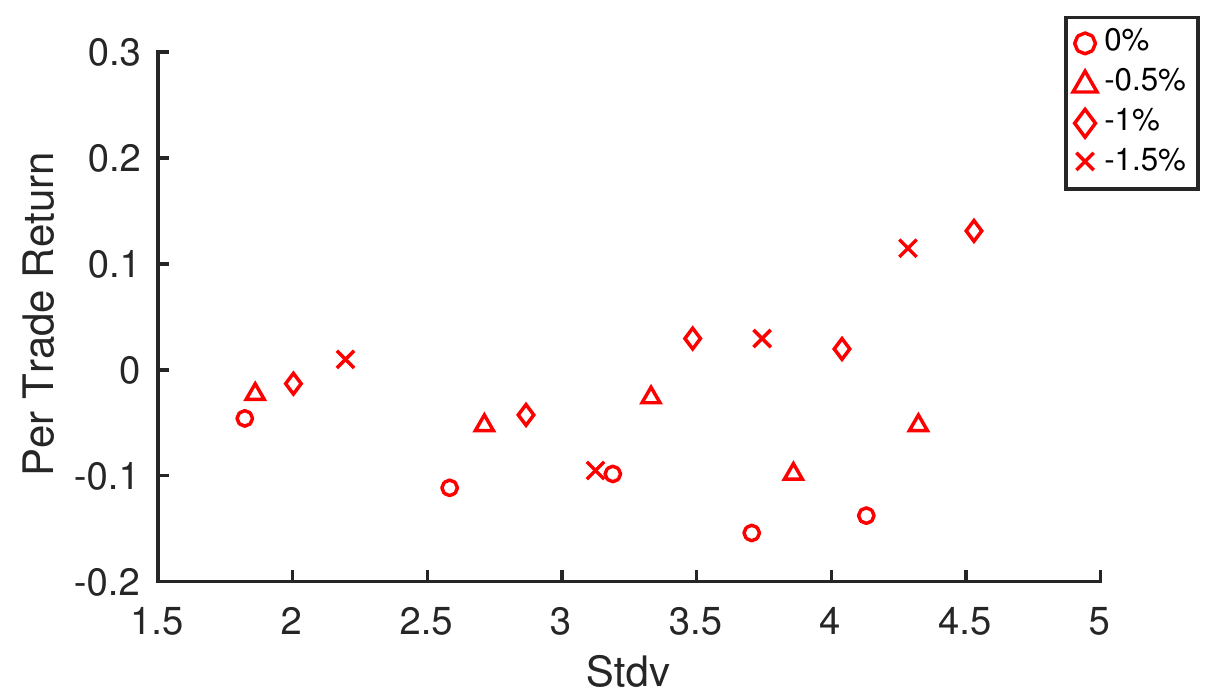}
\caption{ }
\end{subfigure}
\begin{subfigure}[b]{0.5\textwidth}
\includegraphics[width=\textwidth]{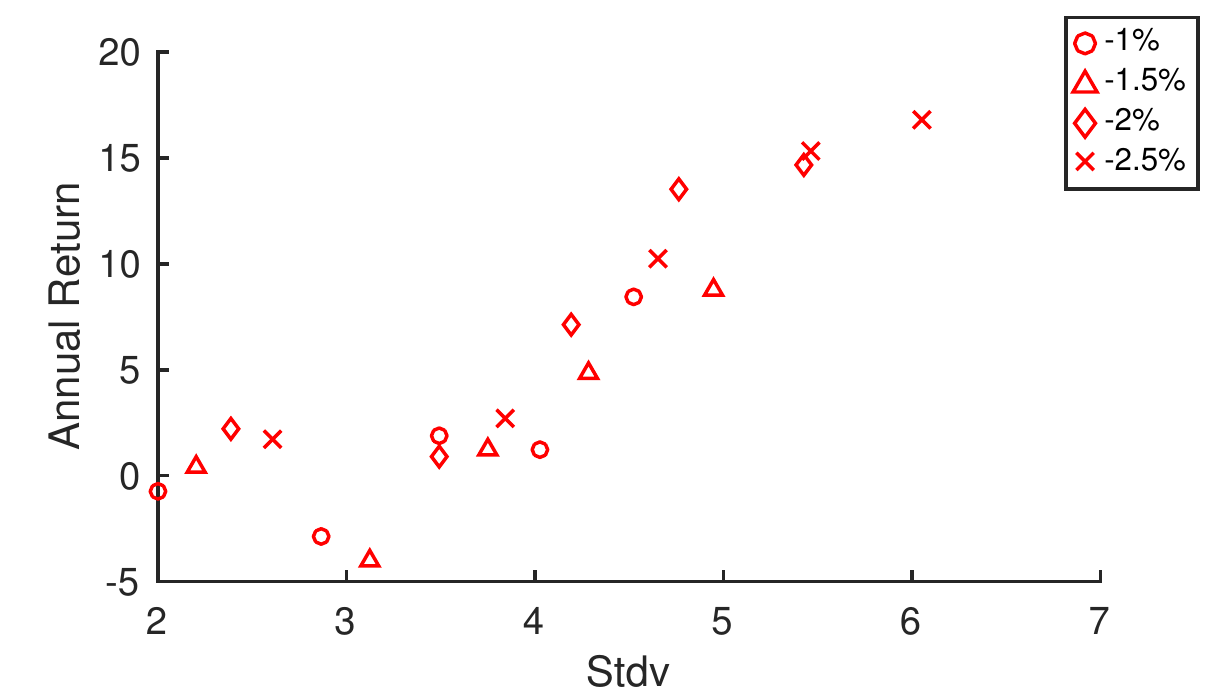}
\caption{ }
\end{subfigure}
\begin{subfigure}[b]{0.5\textwidth}
\includegraphics[width=\textwidth]{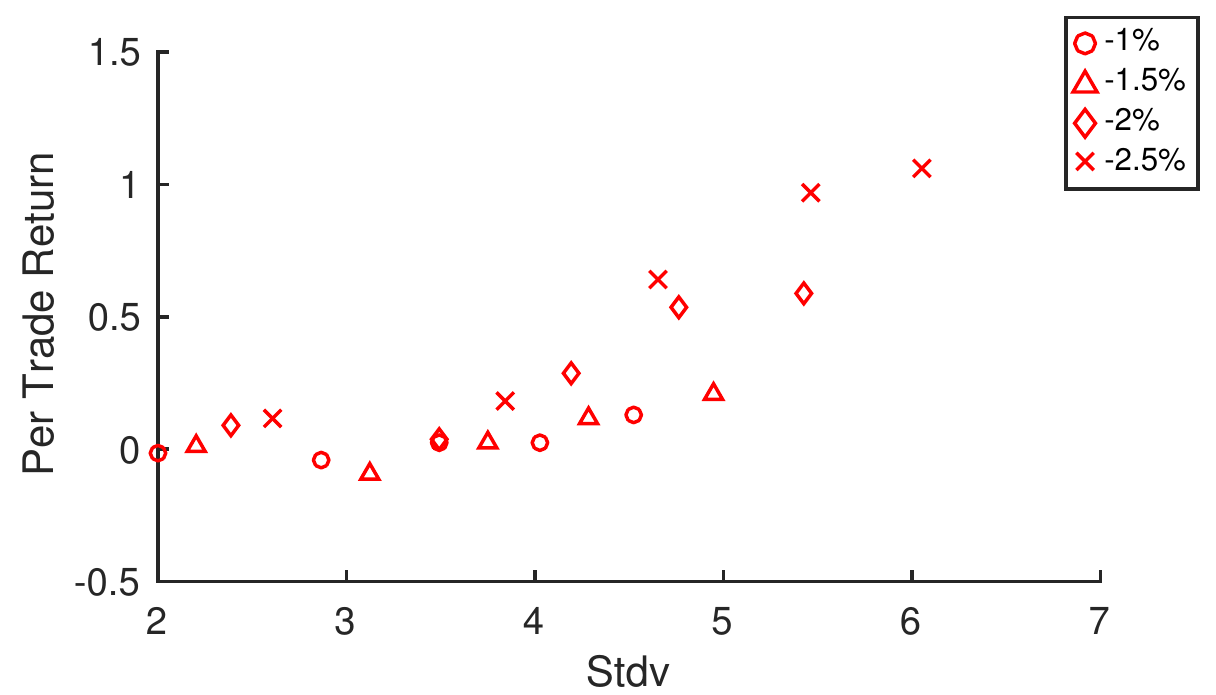}
\caption{ }
\end{subfigure}
\caption{\small{Mean return vs risk (standard deviation)  of the mean-reversion trading strategy for ADR $\in$ $\{$(a),(b) CHL, (c),(d) ASX, (e)-(h) MTU$\}$ based on empirical data from June 15, 2004 to June 13, 2014. }}
\label{fig:trading1}
\end{figure}

\subsection{Applications to foreign equity ETFs}\label{sec4.2}

The proposed mean-reverting pairs trading strategy can be applied to other securities, such as ETFs. Here, we use the ETFs that track equities from Asia, namely China and Japan. This is reasonable since such ETFs give investors exposure to foreign companies as ADRs do at a lower risk due to diversification. Among many foreign ETFs that are currently being actively traded on U.S. markets, we select  FXI (iShares FTSE/Xinhua China 25 Index ETF) and EWJ (iShares MSCI Japan ETF), which are  among the most traded  ETFs of this type.

For this analysis, prices over the 10-year period, 10/26/2005--10/26/2015, are used, and we adopt the same methodology used in Section \ref{sec4.1} with   modification in entry level $k$ values. Specifically, we set $k = \{ -0.4\%, -0.8\%, -1.2\%, -1.6\%\}$, which gives 925, 620, 407, 271 trades respectively for FXI and 819, 436, 217, 108 trades for EWJ. Again, four values of $k$ and five different trade durations $(N)$ from 1 to 5 days are used, creating a total of 20 entry/exit pairs. Profits are normalized to annual and per trade returns using Equation \eqref{eq:profit}. Again, recall that a long position of \$100 in ETF and a short position of \$100 in SPY are entered whenever the return spread, ETF-SPY, below $k\%$ is observed and the positions are closed after N days.

In general, the ETFs also perform well under the pairs trading strategy, generating steadily positive payoffs both annually and per trade. For trades involving  FXI or EWJ,   there is a clear pattern reflecting the tradeoff between risk and  return (see Figure \ref{fig:trading2}).  However, EWJ performs relatively poorly with less negative entry level (especially when $k=-0.4\%$). This can be resolved by properly lowering the entry level.

\section{Conclusion}\label{conclu}
As ADRs have become efficient  financial instruments for investing in foreign companies, it is crucial to understand their return behavior and statistical properties.  With ample empirical data, we have examined the return dynamics of  ADRs, specifically those from Asian countries. Our study has led to a number of new findings. First, ADR returns are shown to be non-normal with heavy two-sided  tails. Second, most ADRs, except TSM and Indian ADRs included in our study, exhibit wider price fluctuations during night when the U.S. market is closed and their originating home markets are open. Third, we quantify the high correlation between the  intraday returns and  the U.S. market returns. Finally,  the return spreads between ADR and SPY, both day and night, are found to be  mean-reverting, which is fitted well to the OU model. This observation is directly utilized in constructing   pairs trading strategies, generating higher annual profits with less volatility, compared to simple by-and-hold strategies for ADRs.  

\begin{figure}[t]
\begin{subfigure}[b]{0.5\textwidth}
\includegraphics[width=\textwidth]{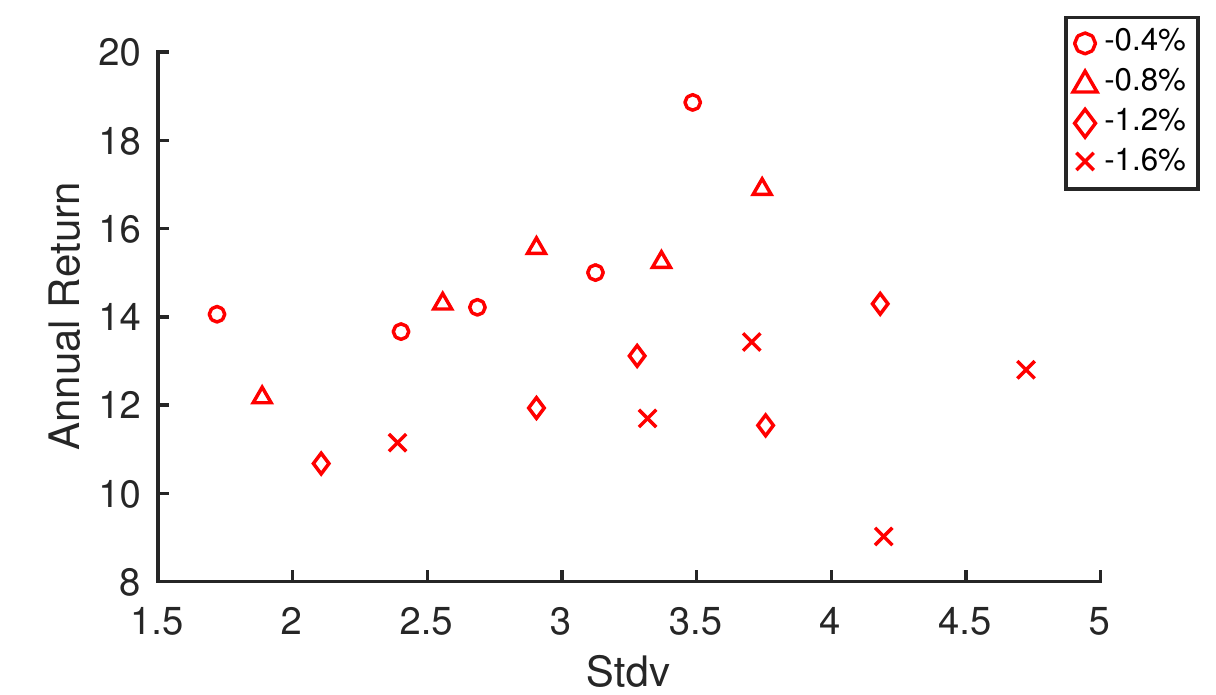}
\caption{ }
\end{subfigure}
\begin{subfigure}[b]{0.5\textwidth}
\includegraphics[width=\textwidth]{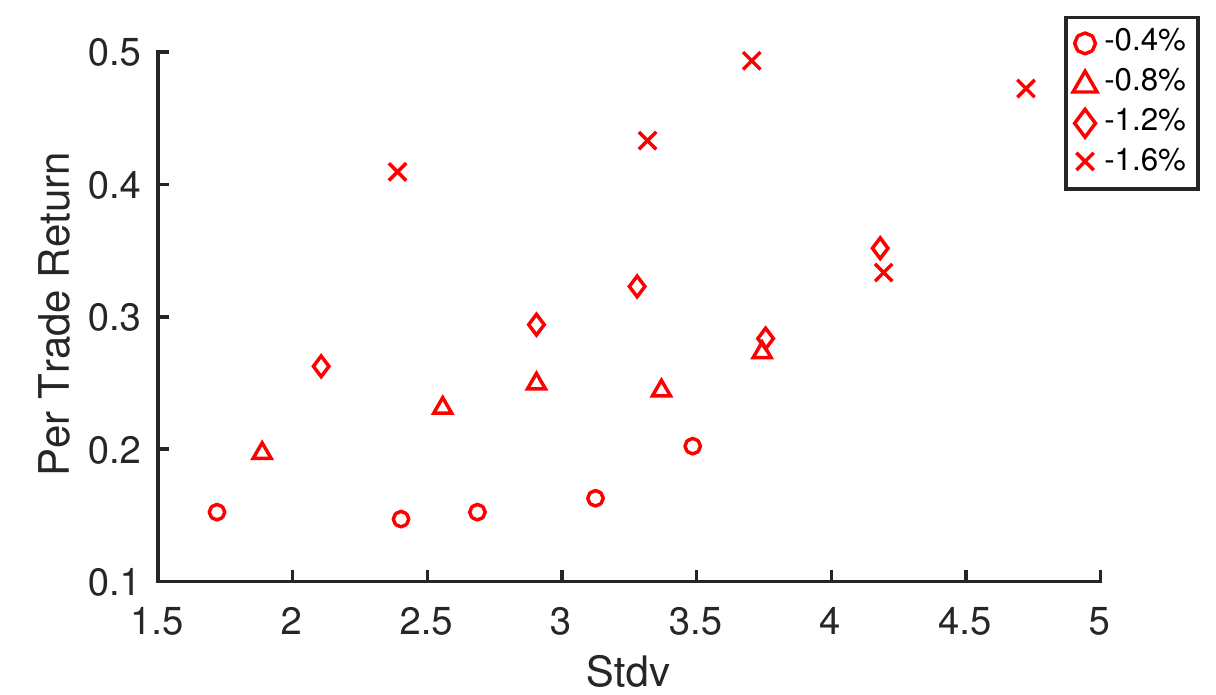}
\caption{ }
\end{subfigure}
\begin{subfigure}[b]{0.5\textwidth}
\includegraphics[width=\textwidth]{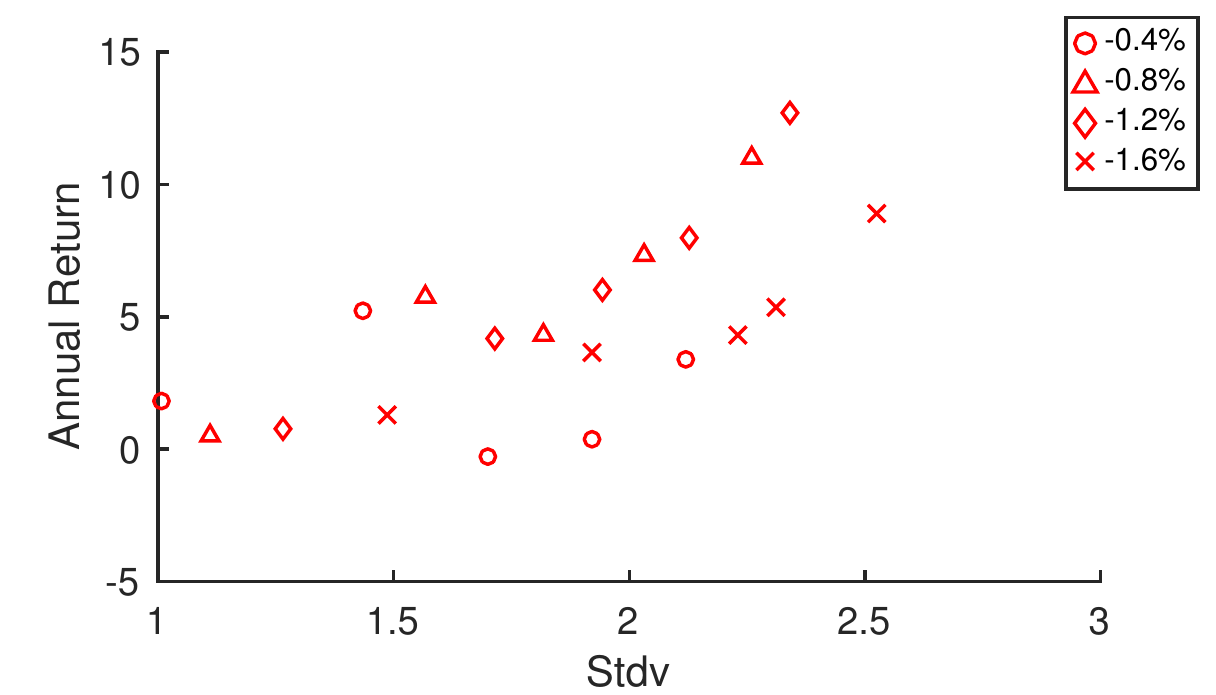}
\caption{ }
\end{subfigure}
\begin{subfigure}[b]{0.5\textwidth}
\includegraphics[width=\textwidth]{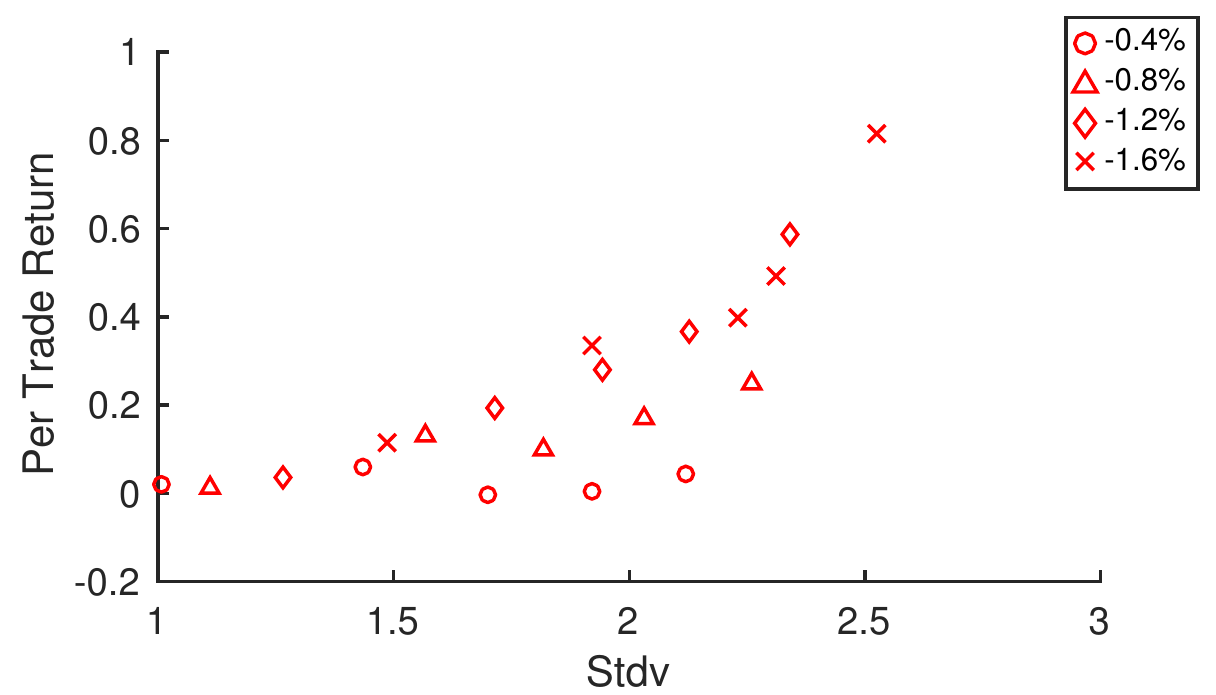}
\caption{ }
\end{subfigure}
\caption{\small{Profitability of  trading FXI ((a) \& (b)) and  EWJ ((c) \& (d)). For each ETF,  mean annual return and mean per trade return are scatter plotted respectively against standard deviation of the return.}}
\label{fig:trading2}
\end{figure}

Although our study is based on Asian ADRs, much of the analytical framework  can be   applied to other domestic and foreign equities, commodity ETFs as well as ETFs with foreign underlying assets, for which pairs trading strategies are common (see e.g. \cite{guoleung}). Also, our threshold-type pairs trading strategies can be implemented on not only a single ADR but also a portfolio of ADRs and/or  equities, constructed for mean-reverting returns (see \cite{LeungLi2015OU} for   illustrative examples). For future research, one direction is to investigate the pricing of options on ADRs in the U.S. and compare to the  options written on the same company stocks listed on local exchanges.

\begin{small}
\bibliography{mybib}

\begin{thebibliography}{}

\bibitem[\protect\astroncite{Bergomi}{2010}]{bergomi2010correlations}
Bergomi, L. (2010).
\newblock Correlations in asynchronous markets.
\newblock {\em Available at SSRN 1635866}.

\bibitem[\protect\astroncite{Cai et~al.}{2011}]{cai2011pricing}
Cai, C.~X., McGuinness, P.~B., and Zhang, Q. (2011).
\newblock The pricing dynamics of cross-listed securities: The case of
  {C}hinese {A}-and {H}-shares.
\newblock {\em Journal of Banking \& Finance}, 35(8):2123--2136.

\bibitem[\protect\astroncite{Eichler et~al.}{2009}]{eichler2009adr}
Eichler, S., Karmann, A., and Maltritz, D. (2009).
\newblock The {ADR} shadow exchange rate as an early warning indicator for
  currency crises.
\newblock {\em Journal of Banking \& Finance}, 33(11):1983--1995.

\bibitem[\protect\astroncite{Guo and Leung}{2015}]{guoleung}
Guo, K. and Leung, T. (2015).
\newblock Understanding the tracking errors of commodity leveraged {ETF}s.
\newblock In Aid, R., Ludkovski, M., and Sircar, R., editors, {\em Commodities,
  Energy and Environmental Finance, Fields Institute Communications}, pages
  39--63. Springer.

\bibitem[\protect\astroncite{Gupta et~al.}{2016}]{gupta2016linkages}
Gupta, R., Yuan, T., and Roca, E. (2016).
\newblock Linkages between the {ADR} market and home country macroeconomic
  fundamentals: Evidence in the context of the {BRIC}s.
\newblock {\em International Review of Financial Analysis}, 45:230--239.

\bibitem[\protect\astroncite{He and Yang}{2012}]{he2012day}
He, H. and Yang, J. (2012).
\newblock Day and night returns of {C}hinese {ADR}s.
\newblock {\em Journal of Banking \& Finance}, 36(10):2795--2803.

\bibitem[\protect\astroncite{Koulakiotis et~al.}{2010}]{koulakiotis2010impact}
Koulakiotis, A., Lyroudi, K., Thomaidis, N., and Papasyriopoulos, N. (2010).
\newblock The impact of cross-listings on the {UK} and the {G}erman stock
  markets.
\newblock {\em Studies in Economics and Finance}, 27(1):4--18.

\bibitem[\protect\astroncite{Lee et~al.}{2015}]{lee2015industry}
Lee, C.-C., Chang, C.-H., and Chen, M.-P. (2015).
\newblock Industry co-movements of {A}merican depository receipts: Evidences
  from the copula approaches.
\newblock {\em Economic Modelling}, 46:301--314.

\bibitem[\protect\astroncite{Leung and Li}{2015}]{LeungLi2015OU}
Leung, T. and Li, X. (2015).
\newblock Optimal mean reversion trading with transaction costs and stop-loss
  exit.
\newblock {\em International Journal of Theoretical \& Applied Finance},
  18(3):15500.

\bibitem[\protect\astroncite{Leung and Li}{2016}]{meanreversionbook2016}
Leung, T. and Li, X. (2016).
\newblock {\em Optimal Mean Reversion Trading: Mathematical Analysis and
  Practical Applications}.
\newblock Modern Trends in Financial Engineering. World Scientific, Singapore.

\bibitem[\protect\astroncite{Leung and Santoli}{2016}]{LeungSantoli2016ETF}
Leung, T. and Santoli, M. (2016).
\newblock {\em Leveraged Exchange-Traded Funds: Price Dynamics and Options
  Valuation}.
\newblock SpringerBriefs in Quantitative Finance, Springer, New York.

\bibitem[\protect\astroncite{Madan}{2012}]{madan}
Madan, D.~B. (2012).
\newblock Jointly modeling the prices of {A}merican depository receipts, the
  local stock and the {US} dollar.
\newblock {\em Journal of Investment Strategies}, 1(4):3--19.

\bibitem[\protect\astroncite{Patel}{2015}]{patel2015adrs}
Patel, S.~A. (2015).
\newblock {ADR}s and underlying stock returns: empirical evidence from {I}ndia.
\newblock {\em AI \& SOCIETY}, 30(2):299--310.

\bibitem[\protect\astroncite{Riedel and Wagner}{2015}]{riedel2015risk}
Riedel, C. and Wagner, N. (2015).
\newblock Is risk higher during non-trading periods? the risk trade-off for
  intraday versus overnight market returns.
\newblock {\em Journal of International Financial Markets, Institutions and
  Money}, 39:53--64.

\bibitem[\protect\astroncite{Rodr{\'\i}guez and
  Toledo}{2015}]{rodriguez2015chinese}
Rodr{\'\i}guez, J. and Toledo, W. (2015).
\newblock C{}hinese single-listed {ADR}s: returns and volatility.
\newblock {\em International Journal of Managerial Finance}, 11(4):480--502.

\bibitem[\protect\astroncite{Williams}{2016}]{williams2016currency}
Williams, O. (2016).
\newblock Foreign currency exposure within country exchange-traded funds.
\newblock {\em Studies in Economics and Finance}, 33(2):222--243.

\end{thebibliography}
\end{small}

\end{document}